\documentclass[11pt]{article}
\usepackage{amsmath,amssymb,color,graphics,epsfig}
\usepackage{graphicx}
\usepackage{subfigure}

\usepackage{amsfonts}

\newcommand{\be}{\begin{equation}}
\newcommand{\ee}{\end{equation}}
\newcommand{\bea}{\setlength\arraycolsep{2pt} \begin{eqnarray}}
\newcommand{\eea}{\end{eqnarray}}
\newcommand{\nn}{\nonumber}
\newcommand{\mm}{\mathrm}
\newcommand{\mc}{\mathcal}

\def\ft#1#2{{\textstyle{\frac{\scriptstyle #1}{\scriptstyle #2} } }}
\def\fft#1#2{{\frac{#1}{#2}}}

\def\0{{\sst{(0)}}}
\def\1{{\sst{(1)}}}
\def\2{{\sst{(2)}}}
\def\3{{\sst{(3)}}}
\def\4{{\sst{(4)}}}
\def\5{{\sst{(5)}}}
\def\6{{\sst{(6)}}}
\def\7{{\sst{(7)}}}
\def\8{{\sst{(8)}}}
\def\sst#1{{\scriptscriptstyle #1}}

\begin{document}

\begin{flushright}
\end{flushright}

\vspace{25pt}
\begin{center}
{\large {\bf On holographic braneworld cosmology }}

\vspace{10pt}
 Zhong-Ying Fan$^1$\\

\vspace{10pt}
$^1${ Department of Astrophysics, School of Physics and Materials Science, \\
 Guangzhou University, Guangzhou 510006, China }\\
\smallskip

\vspace{40pt}

\underline{ABSTRACT}
\end{center}
Recently, S. Antonini and B. Swingle builded a holographic model for braneworld cosmology by introducing an ``end-of-the-world" (ETW) brane moving in a charged black hole spactime. In this paper, we will show that a holographic description of braneworld cosmology is possible for a general black hole spacetime with a pure tension brane if one implements mixed boundary conditions on the ETW brane. As a simple example, we study AdS-Schwarzschild black holes and show that a sensible Euclidean braneworld solution is compatible with localization of gravity on the ETW brane.

\vfill {\footnotesize  Email: fanzhy@gzhu.edu.cn\,.}

\thispagestyle{empty}

\pagebreak

\tableofcontents
\addtocontents{toc}{\protect\setcounter{tocdepth}{2}}





\section{Introduction}

The Anti-de Sitter/Coformal Field Theory (AdS/CFT) correspondence proposes that a $(d+1)$-dimensional quantum gravity theory is dual to a $d$-dimensional quantum filed theory without gravity residing on the asymptotic boundary of bulk spacetime
\cite{Maldacena:1997re,Witten:1998qj,Gubser:1998bc}. It provides a new approach to study both quantum gravity and quantum field theories at strong coupling limit, becoming an exciting research area in recent two decades. It is natural to ask whether cosmological physics can be studied in this framework. Some earlier works on holographic cosmology can be found in literature, for example \cite{Verlinde:2000wg,Savonije:2001nd,Gregory:2002am,Bilic:2015uol,Bernardo:2018cow,McFadden:2009fg}, where a conformal field theory (CFT) lives on the brane and is coupled to gravity.

Recently, it was established \cite{Antonini:2019qkt} that a non-perturbative CFT description of Friedmann-Lema\^{i}tre-Robertson-Walker (FLRW) cosmology can be realized in the Anti-de Sitter/Boundary Conformal Field Theory (AdS/BCFT) correspondence
\cite{Takayanagi:2011zk,Fujita:2011fp}. The basic idea is relating the excited pure states of the CFT to a two-sided black hole geometry with a dynamical end-of-world (ETW) brane cutting off the left (or right) region of the black hole and replacing the asymptotic boundary therein \cite{Cooper:2018cmb}. It is known that a two-sided wormhole geometry describes an entangled states of two CFTs living in the left and right asymptotic boundary respectively. Now the presence of the ETW brane was proposed to measure the one CFT (for example the left one), destroying the entanglement and hence leaving a pure state for the remaining CFT.

It is intriguing that the motion of the ETW brane could be interpreted as evolution of a FLRW universe for local observers living on the brane, where the bulk radial coordinate plays the role of the scale factor in cosmology \cite{Binetruy:1999ut,Binetruy:1999hy,Padilla:2001jz,Padilla:2002tg}. Hence, this model provides a holographic description for braneworld cosmology using the observables in the dual CFT associated to the right asymptotic region of black holes. This is significantly different from the earlier works \cite{Verlinde:2000wg,Savonije:2001nd,Gregory:2002am,Bilic:2015uol,Bernardo:2018cow,McFadden:2009fg}, where the CFT lives on the brane.

However, there are two things that should be considered carefully. The first is the dual CFT should be well defined. This demands that the preparation time of the CFT, defined in the Euclidean path integral, is positive definite \cite{Cooper:2018cmb}. The second is if the bulk is $5$-dimensional, for observers living on the brane, gravity should be effectively $4$-dimensional and is locally localized on the brane. In other words, ordinary gravity is observed only on non-cosmological scales and in a limited range of times.

It was shown in \cite{Antonini:2019qkt} that for AdS Reissner-Nordstr\"{o}m (RN) black holes with a pure tension brane, the above two conditions are compatible by implementing Neumann boundary condition on the brane. It provides a first proof-of-principle result for holographic braneworld cosmology. However, the above two conditions are not always compatible. As pointed out in \cite{Antonini:2019qkt}, the result there is not attainable in the simplest setup of an AdS-Schwarzschild black hole with a pure tension brane. As a matter of fact, the discussions in \cite{Antonini:2019qkt} are limited to charged black holes and cannot be extended to general black hole backgrounds. This is unsatisfactory from the viewpoint of holographic principle.

A possible remedy of the issue is by introducing localized matter on the ETW brane. However, this is of great difficult in practical applications (there are also many other issues). Instead, in this paper, we will show that the issue can be nicely resolved by imposing a mixed boundary condition on the ETW brane. We prove this for AdS-Schwarzschild black holes. The results are easily generalised to different black hole backgrounds. We thus illustrate a simple setup for braneworld cosmology in holography.

The paper is organized as follows. In section 2, we briefly review the AdS/BCFT correspondence and the role of boundary conditions. In section 3, we study the motion of the ETW brane in Euclidean AdS-Schwarzschild black holes and show that sensible Euclidean solutions can be obtained by imposing a mixed boundary condition on the brane. In section 4, we study the frequency of the quasi bound mode for large black holes in adiabatic approximation. We find that gravity is locally localized on the brane for parameters compatible with sensible Euclidean solutions. We conclude in section 5.

\section{AdS/BCFT }
Boundary conditions play an essential role in the formulation of AdS/BCFT correspondence. Consider a CFT living in a $d$-dimensional manifold $M$ with a $(d-1)$-dimensional boundary $P$. The bulk is a $(d+1)$-dimensional AdS space $N$, which is anchored on $M$ and has an extra boundary $Q$ in the deep interior, with $\partial Q=\partial M=P$. In general, for a pure tension brane $Q$, the gravitational action is given by
\be I_L=\fft{1}{16\pi G}\int_N \sqrt{-g} \,\big( R-2\Lambda \big)+\fft{1}{8\pi G}\int_M \sqrt{-\gamma}\, K+\fft{1}{8\pi G}\int_Q \sqrt{-h}\, \big(K-(d-1)T \big)+\fft{1}{8\pi G}\int_P \,\sqrt{\sigma}\, a \,,\ee
where $a=\mm{arccos}\big(n_M\cdot n_Q \big)$ is the corner term associated to the codimension-2 boundary $P$ (from the bulk point of view). This term is introduced for well-defined variational principle in the presence of non-smooth boundaries \cite{Hayward:1993my}. Otherwise, it can be dropped in the smooth limit, where variation of the action yields\footnote{Dirichlet boundary condition was imposed on the asymptotic boundary $M$.}
\be \delta I_L=\int_Q \sqrt{-h}\Big( K_{ab}-\big(K-(d-1)T \big)h_{ab}\Big)\delta h^{ab}\,.\ee
Clearly, to have dynamics on $Q$, one should introduce Neumann boundary condition
\be\label{nbc}  K_{ab}-\big(K-(d-1)T \big)h_{ab} =0\,,\ee
 which allows the brane moving (expanding or contracting) dynamically. However, this case has many problems, as pointed out in \cite{Miao:2017gyt,Chu:2017aab}. On one hand, since $Q$ is a codimension-$1$ hypersurface in the bulk, it should be characterized by a single embedding function while the Neumann boundary condition (\ref{nbc}) is a tensorial equation, which generally does not give a self-consistent solution except some simple cases \cite{Takayanagi:2011zk,Fujita:2011fp}. On the other hand, the AdS/BCFT correspondence should apply to general shape of the boundary $M$. A possible remedy of the issue is by introducing localized matter on the brane $Q$ so that the equation (\ref{nbc}) (with stress tensor of matter fields introduced on the r.h.s) can be solved consistently for a general shaped asymptotic boundary. However, this approach is of great difficult in practical applications and cannot be tested easily. As a consequence, whether it provides a consistent story for the boundary theory remains open.

In \cite{Miao:2017gyt,Chu:2017aab}, the authors provide another solution to the issue: imposing a mixed boundary condition on $Q$
\be K=d T\,.\label{mixbc}\ee
This does not contradict with the variational principle since in general the corner $P$ is non-smooth and the bulk action should pick up a joint term. Moreover, despite that the condition looks simple, it produced correct holographic boundary Weyl anomaly for the dual CFT. This is highly nontrivial. It partly establishes that the above mixed boundary condition is physically sensible in studying holographic dual of CFTs with boundaries.

Return to holographic description of braneworld cosmology. The bulk interior boundary $Q$ corresponds to an end-of-the-world (ETW) brane, which replaces the left/right asymptotic boundary of a two-sided black hole. In \cite{Antonini:2019qkt}, the authors first show that the AdS Reissner-Nordstr\"{o}m (RN) black holes with a pure tension brane are Euclidean sensible for proper parameters by imposing Neumann boundary condition on the ETW brane\footnote{They also show that for a given tension and a positive preparation time, large charged black holes are dominate in the Euclidean path integral of gravity.}. They further show that for the same parameters, gravity is locally localized on the brane. This gives a first example to holographic description of cosmology. However, they mispointed out that the mixed boundary condition leads to the same results as the Neumann boundary condition. As a consequence, they claimed that the above results are not attainable in AdS-Schwarzschild black holes with a pure tension brane because a sensible Euclidean solution mutually excluded with localization of gravity on the brane.

In this paper, we will show that the above mixed boundary condition gives new solutions to motion of the ETW brane, which was characterized by one integration constant $q$, referred to as {\it dark charge}. This is a general feature of the boundary condition, which does not rely on the black hole background. As a simple example, we show that by tuning the {\it dark charge} $q$, one can obtain both a sensible Euclidean solution and gravity localization on the brane for AdS-Schwarzschild black holes with a pure tension brane.

\section{The end-of-the-world brane in AdS}

Consider a two-sided static spherically symmetric AdS black hole of the form
\be ds^2=-f(r)dt^2+\fft{dr^2}{f(r)}+r^2 d\Omega_{d-1}^2 \,,\ee
and introduce an end-of-the-world (ETW) brane (spherically symmetric) on the left side. The wormhole geometry with a ETW brane replacing the left asymptotic boundary is dual to an excited pure state of the CFT living in the right asymptotic boundary. The motion of the brane is characterized by $r=r(t)$, which encodes the evolution of FLRW universe.

However, a faithful holographic description for braneworld cosmology demands that the dual CFT state is well defined, at least in the Hilbert space of states. Let us turn to analyze Euclidean CFT, which is dual to Euclidean black holes, using imaginary time $\tau=it$. The boundary state will evolve for an finite amount of time $\tau_0$, referred to as preparation time
\be\label{bdystate} |\Psi\rangle=e^{-\tau_0 H}|B\rangle \,,\ee
where the ETW brane can be viewed as bulk extension of the past boundary of the CFT, corresponding to the past boundary condition in the Euclidean path integral. Clearly, the preparation time, which is a property of the CFT state, should be positive definite. Otherwise, the state will be non-normalizable and hence is not in the Hilbert space.

\begin{figure}
  \centering
  \includegraphics[width=200pt]{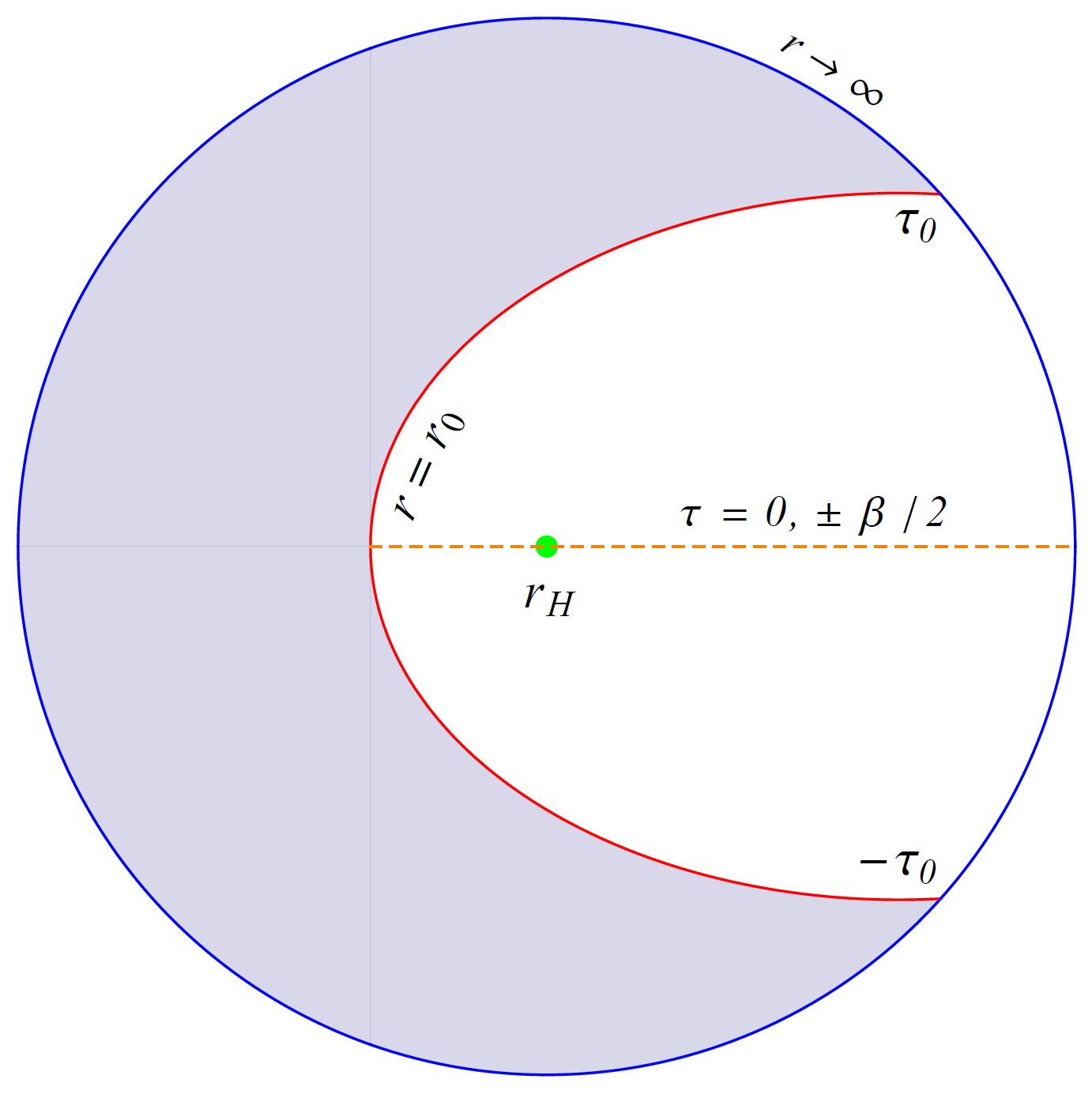}
  \caption{Euclidean brane trajectory (red). The brane contracts from the asymptotic boundary at $\tau=-\tau_0$, reaches a minimal radius $r=r_0$ and then returns to the asymptotic boundary again at $\tau=\tau_0$. The shaded region is excised by the brane.}
\label{trajectory}\end{figure}

In the bulk, the total range of the Euclidean time $\tau$ is given by the the inverse temperature $\beta$ of the black hole. However, for a given preparation time $\tau_0>0$, the brane trajectory excises part of this total range of time: it begins to contract from the asymptotic boundary at $\tau=-\tau_0$, reaches a minimal radius $r=r_0$ and then returns to the asymptotic boundary again at $\tau=\tau_0$, see Fig. \ref{trajectory}. Hence, the CFT state is defined only in the interval $\tau\in [-\tau_0\,,\tau_0]$. The preparation time can be evaluated as
\be \tau_0=\beta/2-\Delta \tau\,, \ee
where $\Delta \tau$ is the Euclidean time associated to half trajectory of the brane.

\subsection{Euclidean analysis}\label{sec31}
Let us study the motion of the ETW brane and calculate the preparation time explicitly. The unit dual normal vector of the brane is given by
\be\label{norm} n_\mu=N(-r'\,,1\,,\mathbf{0})\,,\qquad N=\sqrt{\fft{f(r)}{f^2(r)+(r')^2}} \,,\ee
where a prime denotes the derivative with respect to $\tau$. The induced metric on the brane turns out to be
\be ds^2_{\mm{ETW}}\equiv h_{ab}dx^a dx^b=N^{-2}d\tau^2+r^2 d\Omega_{d-1}^2 \,.\ee
The extrinsic curvature can be calculated from the definition $K_{ab}=e^\mu_a e^\nu_b\nabla_\mu n_\nu$, where $e^\mu_a=dx^\mu/dy^a$ and the embedding function is $x^\mu=x^\mu(y^a)$. With $e^\mu_\tau=(1\,,r'\,,\mathbf{0})$, one finds
\be K_{\tau\tau}=N\Big(-r''+\fft{3f_r(r)}{2f(r)}(r')^2+\fft{f(r)f_r(r)}{2} \Big)\,,\qquad K_{ij}=\fft{N f(r)}{r}h_{ij} \,,\ee
where $f_r=df/dr$ and $i\,,j$ labels the angular coordinates. According to the mixed boundary condition (\ref{mixbc}), one finds\footnote{There are a variety of choices for the mixed boundary condition imposed on the brane, for example
\be C^{ab}\Big[ K_{ab}-\big(K-(d-1)T \big)h_{ab} \Big]=0\,, \ee
where $C^{ab}$ is a projector. In the main text, we study the simplest case $C_{ab}=h_{ab}$. Different choices of the boundary condition is discussed in the Appendix A.   }
\be r''-\fft{3f_r(r)}{2f(r)}(r')^2-\fft{f(r)f_r(r)}{2}-\fft{(d-1)f(r)}{N^2 r}+\fft{dT}{N^3}=0\,. \ee
This is a second order nonlinear ODE, describing the motion of the brane. However, using the following relations
\be r'=\pm\sqrt{\fft{f(r)}{N^2}-f^2(r)}\,,\qquad r''=\fft12 \partial_r\Big( \fft{f(r)}{N^2}-f^2(r)\Big) \,,\ee
the above equation simplifies to
\be 0=\sqrt{g}\,\big(K-dT \big)=\partial_r \big(f(r)r^{d-1} N-T r^d\big)\,.\ee
It leads to a first integral
\be f(r)r^{d-1}N-T r^d=q^{d-1} \,,\ee
where $q$ is an integration constant, referred to as {\it dark charge}. We deduce

\be \fft{dr}{d\tau}=\pm \fft{f(r)}{\varphi(r)}\sqrt{f(r)-\varphi(r)^2}\,,\qquad \varphi(r)=Tr+\big(\fft{q}{r}\big)^{d-1} \,,\label{firstint}\ee
where the $\pm$ sign corresponds to expansion or contraction of the brane ($\tau$ is an increasing or decreasing function of $r$). We are aware of that the above solution includes the result of Neumann boundary condition \cite{Antonini:2019qkt} as a special case $q=0$. However, to have a sensible Euclidean solution, one needs a very large $q$: $q/r_H\gg 1$, as will be shown later. This partly explains why the results in \cite{Antonini:2019qkt} are not attainable for AdS-Schwarzschild black holes.

The inversion point of the ETW brane $r_0$ is determined by
\be\label{inversion} f(r_0)=\varphi^2(r_0) \,,\ee
where $r_0$ stands for the minimal real root of the equation. Gravity localization on the brane demands that the brane should be far away from the event horizon: $r_0/r_H\gg 1$. Otherwise, gravitons will leak into the bulk, fall into the horizon and hence delocalize the gravity. We may take $r_0$ close to the asymptotic AdS boundary such that $\varphi(r_0)\rightarrow T r_0$. This implies $T^2< 1/L^2$, where $L$ is the AdS radius. We shall focus on a positive tension brane $0<T< 1/L$ since physical meaning of the solutions with a negative tension brane are unclear \cite{Antonini:2019qkt,Cooper:2018cmb}.

\begin{figure}
  \centering
  \includegraphics[width=250pt]{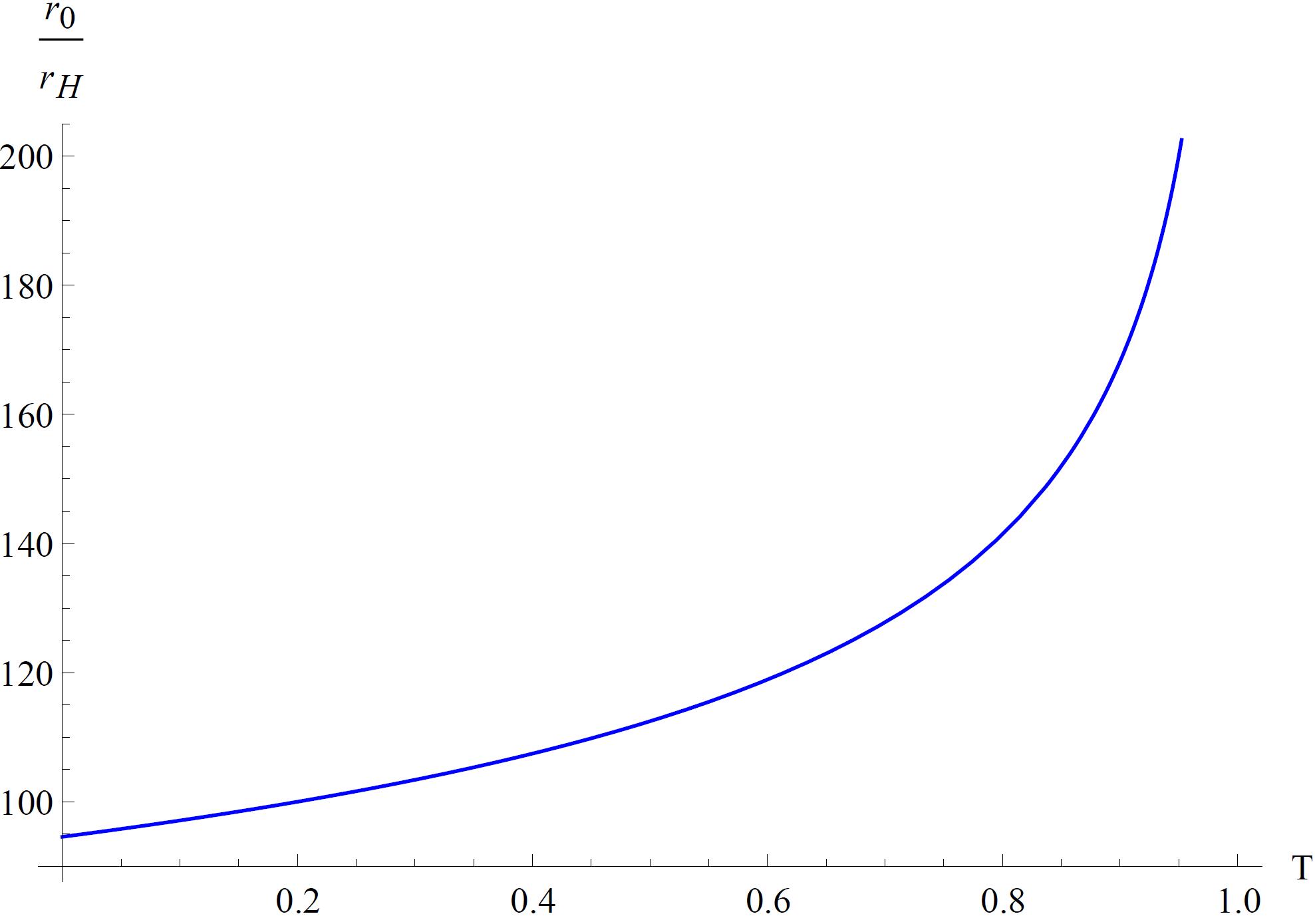}
  \caption{The minimal radius $r_0$ is an increasing function of the brane tension $T$. We have set $L=1$, $r_H=100\,,q=200000$. }\label{tension}
\end{figure}

In the following, we consider the simplest case: AdS-Schwarzschild black holes
\be f(r)=\fft{r^2}{L^2}+1-\fft{2\mu}{r^{d-2}} \,,\ee
where $\mu$ is the mass parameter of the black hole (the black hole mass is $M=\fft{(d-1)\Omega_{d-1}}{8\pi G}\,\mu$).
In Fig. \ref{tension}, we plot the ratio $r_0/r_H$ as a function of the brane tension. The brane with a near critical tension $T\rightarrow T_{crit}=1/L$ can have a very large ratio $r_0/r_H$ by increasing the dark charge $q$. For example, for $r_H=100\,,T=0.99999\,,q=200000$, $r_0/r_H\sim 200$. For larger black holes with the same tension brane, the dark charge will be larger to have the same ratio. This is consistent with naive expectations since gravity localization becomes more difficult for larger black holes and hence needs a larger $q$.

\begin{figure}
  \centering
  \includegraphics[width=215pt]{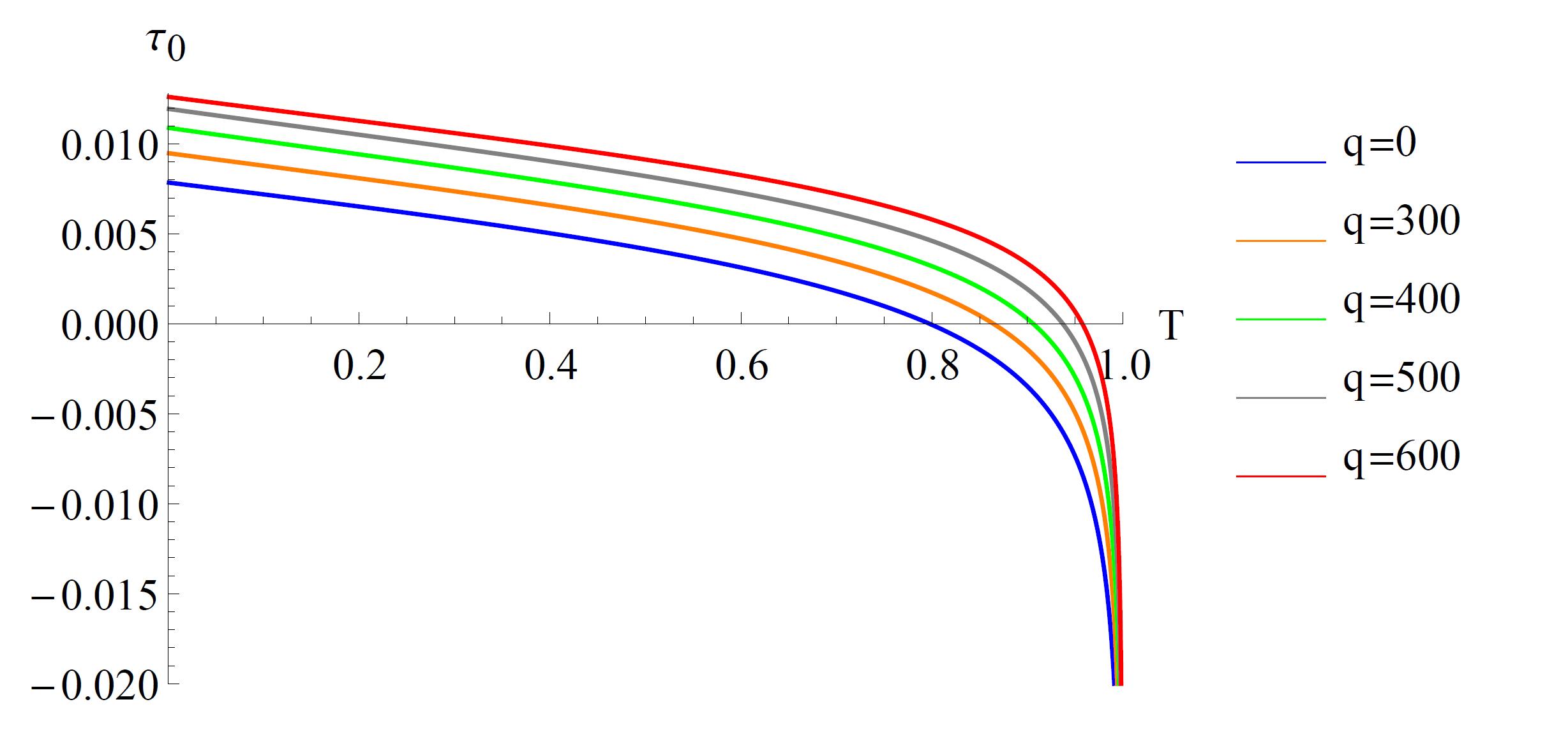}
  \includegraphics[width=215pt]{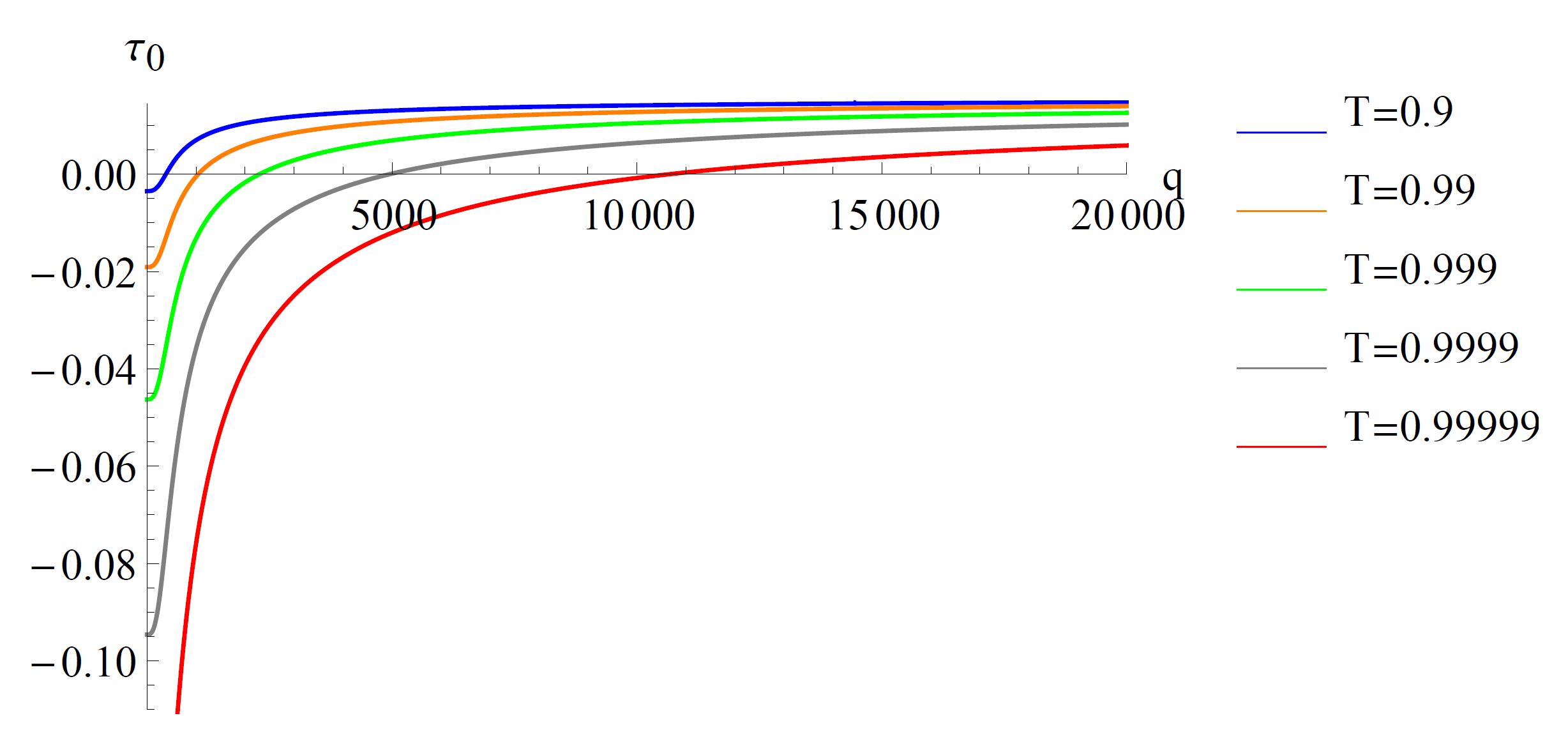}\\
  \caption{The preparation time $\tau_0$ as a function of $T$ (left panel) or $q$ (right panel) for a large black hole with $r_H=100$. }\label{tau}
\end{figure}

Explicitly, the preparation time is given by
\be \tau_0=\fft{\beta}{2}-\int_{r_0}^\infty dr\, \fft{\varphi( r)}{f( r)\sqrt{f( r)-\varphi^2( r)}}  \,.\ee
The numerical results are shown in Fig. \ref{tau}. For a given $q$, the preparation time is a decreasing function of the brane tension, and is positive definite for a non-critical brane, see the left panel. On the other hand, for a near critical brane, a positive $\tau_0$ demands a sufficiently large $q$, depending on the black hole sizes, as shown in the right panel.

It is worth emphasizing that one can always find parameters $(T\,,r_H\,,q)$ allowing both a large ratio $r_0/r_H\gg 1$ and a positive preparation time. We may expect that for AdS-Schwarzschild black holes, the boundary CFT state is well defined and is consistent with locally localization of gravity on the ETW brane. We will show this explicitly in section 4.

Last but not least, we show in the Appendix \ref{actioncompare} that for a positive preparation time $\tau_0$, the large black hole solution is the dominant phase, in the thermodynamic ensemble with a fixed dark charge $q$.



\subsection{Lorentian analysis}

In real time, the CFT state can be obtained by analytically continuing $\tau\rightarrow it$ in Eq.(\ref{bdystate}). Correspondingly, the bulk geometry is the maximally extended AdS-Schwarzschild black holes with the ETW brane cutting off part of the left asymptotic region. However, changing to Lorentzian signature will flip the sign of the square root in Eq.(\ref{firstint}). It implies that the minimum radius in Euclidean signature now becomes the maximal one. Hence, the brane contracts/expands in the bulk interior $r\leq r_0$, crossing the horizon and reaching/leaving the singularity. This is the physical picture relevant to braneworld cosmology.

Defining the proper time $d\lambda^2=\Big(f(r)-\fft{r'^2(t)}{f(r)} \Big)dt^2$, the reduced metric on the brane can be written as
\be ds^2_{\mm{ETW}}=-d\lambda^2+r^2(\lambda)d\Omega^2_{d-1} \,.\ee
 It corresponds to a FLRW universe with curvature $K=0$. The bulk radial coordinate plays the role of the scale factor in cosmology. From the motion of the brane, it is straightforward to derive the Friedmann equation. For Schwarzschild black holes, one finds
\be \Big( \fft{\dot{r}}{r}\Big)^2=\Big(T^2-1/L^2\Big)-\fft{1}{r^2}+\fft{2\big(\mu+T q^{d-1} \big)}{r^d}+\fft{q^{2(d-1)}}{r^{2d}} \,,\ee
where a dot denotes the derivative with respect to the proper time. It implies that comoving observer interprets the motion of the brane as expansion or contraction of a Big Bang-Big Crunch FLRW cosmology. In $d=4$ dimension, this describes a radiation dominated universe with a (very small) negative cosmological constant $3(T^2-1/L^2)$ and energy density proportional to the black hole mass plus the contribution from the dark charge. It is remarkable that the dark charge $q$ gives an extra contribution to the energy density which is naively not visible to bulk observers. However, observers on the brane can definitely find this part of energy via cosmological measurements (This can be said even more precisely by calculating the energy measured by observers on the brane directly \cite{Binetruy:1999ut,Binetruy:1999hy,Padilla:2001jz,Padilla:2002tg}. However, we prefer explaining the energy terms in the Friedmann equation using the stress tensor in the boundary CFT, see the next subsection).

In the AdS/CFT correspondence, black hole mass is usually interpreted as total energy in the boundary theory. However, the above Friedmann equation implies that this may not be true in braneworld cosmology. The black hole mass is only a part of the total energy driving the cosmological evolution. This inspires us that if one interprets the black hole mass as visible matter in our universe, the dark charge may be identified with dark matter in FLRW cosmology. However, this holographic argument neither gives any hint on microstructures of dark matter nor explains its gravitational effects in galactic scales.

Notice that the last term in the Friedmann equation is unconventional. It was not found in earlier literature on braneworld cosmology. However, this term disappears for a different mixed boundary condition. In this case, the dark charge plays completely the same role of black hole mass, see the Appendix A.


\subsection{Boundary stress tensor}
To examine the physical meaning of the dark charge in the boundary CFT, we would like to evaluate the stress tensor in the boundary in Euclidean signature. As usual, we needs renormalizing the gravitational action by introducing proper counter terms. In the $d=4$ dimensional case, the renormalized action is given by
\bea
I_{ren}&=&-\fft{1}{16\pi g}\int_N \sqrt{g}\,\big( R-2\Lambda \big)-\fft{1}{8\pi G}\int_{ETW} \sqrt{h}\,\big(K-(d-1)T \big) \nn\\
       &&-\fft{1}{8\pi G}\int_M \sqrt{\gamma}\,\big( K_M-3-\fft14 R_M\big)-\fft{1}{8\pi G}\int_P \sqrt{\sigma}\,\big( \theta-\theta_0-\fft12 K_P-\alpha R_P\big)\,,\eea
where $K_M\,,R_M$ are extrinsic and intrinsic curvatures of the asymptotic boundary $M$ respectively. $K_P$ is the extrinsic curvature of $P$, which is introduced as the Gibbons-Hawking-York (GHY) surface term for $R_M$. $\theta_0=\theta|_{r\rightarrow \infty}$ and $\alpha=\fft{T}{4\sqrt{1-T^2}}$. Variation of the action yields
\be\label{renaction} \delta I_{ren}=\int_N E.O.M-\fft12 \int_M T^{M}_{\mu\nu}\delta \gamma^{\mu\nu}
-\fft12 \int_{Q} T^{Q}_{\mu\nu}\delta h^{\mu\nu}-\fft12 \int_P T^{P}_{ab}\delta \sigma^{ab} \,,\ee
where
\bea
&&T^{M}_{\mu\nu}=\fft{1}{8\pi G}\Big[K^M_{\mu\nu}-\big( K_M-3\big)g^{M}_{\mu\nu}-\fft12 \mc{G}^{M}_{\mu\nu} \Big]   \,,\nn\\
&&T^{Q}_{\mu\nu}=\fft{1}{8\pi G}\Big[ K_{\mu\nu}-\big(K-3T \big)g_{\mu\nu}  \Big]   \,,\nn\\
&&T^{P}_{ab}=-\fft{1}{8\pi G}\Big[ \fft12\big(K^P_{ab}-K_P \sigma_{ab}\big)+\big(\theta-\theta_0 \big)\sigma_{ab}+2\alpha\, \mc{G}_{ab}^P  \Big]    \,,
\eea
where $\mc{G}^M_{\mu\nu}\,,\mc{G}^P_{ab}$ are the Einstein tensor of $M$ and $P$ respectively. The result for $T^{M}_{\mu\nu}$ is the usual holographic stress tensor of the dual CFT living on the manifold $M$ without extra boundaries. However, now the manifold $M$ has a past boundary and generally there exists a nonzero stress tensor on $P$, depending on the boundary conditions.

Evaluation of the stress tensors on $Q$ yields ( we set $G=1/8\pi$)
\be
T_{\tau\tau}^Q=-\fft{3q^3}{r^4}\fft{f^2(r)}{\phi^2(r)}\,,\qquad T_{ab}^Q=\fft{q^3}{r^2}\,\zeta_{ab}\,,
\ee
where $\zeta_{ab}$ is the metric of a unit $\mathbb{S}^3$. It is clear that for $q=0$, corresponding to the Neumann boundary condition, there exists no energy flux on the ETW brane, as expected. On the other hand, on the past boundary $P$, the vev of the stress tensor can be read off as
\be \langle T_{ab}^P\rangle=\lim_{r\rightarrow \infty}r^2 T_{ab}^P=\Big(\ft{T(3-2T^2)}{8(1-T^2)^{3/2}}+\ft{q^3+\mu T}{\sqrt{1-T^2}} \Big)\zeta_{ab}\,.\ee
It implies that the mixed boundary condition imposed on the ETW brane will increase the stress tensor $\langle T^P_{ab}\rangle$ on the past boundary $P$. In particular, the dark charge $q$ plays the same role of the black hole mass, similar to the situation in the Friedmann equation, despite that it never contributes to the bulk stress tensor $\langle T_{\mu\nu}^M\rangle$. This supports our previous argument in part.


\section{Gravity localization}

We are left to show gravity localization: whether observers on the brane see approximately four dimensional gravity. In the Randall-Sundrum II model \cite{Randall:1999vf}, four dimensional gravity is reproduced on a Minkowski brane embedded in a warped AdS$_5$ spacetime due to the graviton zero-mode that is bound on the brane. The bound mode is lost in the presence of a black hole in the bulk \cite{Seahra:2005us}. Instead, a resonant quasi bound mode persists if the brane is static and far from the black hole horizon \cite{Clarkson:2005mg}. This quasi bound mode is meta-stable because of a finite life time. Hence, localization of gravity on the brane is a local effect, valid to spatial and time scales smaller than the cosmological ones.

In \cite{Clarkson:2005mg}, the authors focused on small AdS Schwarzschild black holes with $r_H<L$. In this case, there exists a potential well for graviton wave function outside the event horizon (see Fig. \ref{potential}), leading to a discrete set of quasi bound modes, referred to as {\it overtones}. On the other hand, for large black holes $r_H>L$, the potential well disappears and there is only a fundamental quasi bound mode. It was argued in \cite{Clarkson:2005mg} that this fundamental mode approaches to the RS II bound mode when the brane is close to asymptotic infinity, irrespective of the black hole sizes. That is in this limit the imaginary part of the frequency of the quasi bound mode vanishes whereas its real part tends to the frequency $\omega_{GR}=\sqrt{\ell(\ell+2)f(y_b)}/y_b$, expected for a four-dimensional gravitational perturbation of the Einstein static universe.

 However, this topic deserves further investigations in current paper. There are two reasons. The first is the brane is static in \cite{Clarkson:2005mg} while in our case it moves dynamically. It is highly nontrivial to implement a proper boundary condition in this case. The second is the above limiting behavior of the fundamental quasi bound mode was criticized in \cite{Antonini:2019qkt}, in which the authors found that when the brane is sufficiently far from the black hole horizon, the quasi bound mode becomes very short-lived and hence gravity localization is lost. We will re-examine this issue using both numerical and half-analytical methods.

\subsection{Master equation and boundary condition}
Consider a linear perturbation around a AdS-Schwarzschild black hole
$\delta g_{\mu\nu}=g_{\mu\nu}-g^{(0)}_{\mu\nu}$. The perturbation can be decomposed into scalar, vector and tensor components \cite{Kodama:2003jz}. The latter corresponds to graviton mode of interest which has
$ \delta g_{r\mu}=0=\delta g_{t\mu}$. As usual, we work in transverse-traceless gauge $\delta g^{\mu}_{\,\,\,\,\mu}=0=\nabla^\mu \delta g_{\mu\nu}$. Then the graviton mode can be expanded as ($ i\,,j=1\,,2\,,\cdots\,,d-1$)
\be \delta g_{ij}(t\,,r\,,x^i)=\sum_{\omega\,,k}r^{\fft{5-d}{2}}\psi(r)e^{i\omega t}\,Y_{ij}^{(k)}(x^i) \,,\ee
where $Y_{ij}^{(k)}$ stands for hyperspherical harmonics, satisfying $\Delta_{d-1} Y_{ij}^{(k)} =-k^2 Y_{ij}^{(k)}$, where $\Delta_{d-1}$ is the Laplacian operator on the unit $(d-1)$-sphere and $k^2=\ell(\ell+d-2)-2$, with $\ell=1\,,2\,,\cdots$ is the generalized angular momentum.

It was established in \cite{Kodama:2003jz} that using the above expansion, the linearized Einstein equation for each tensor mode of the perturbations can be recast into a one-dimensional Schr\"{o}dinger equation for $\psi(r)$. For later convenience, we work directly with adimensional coordinates and parameters. We introduce $y=r/r_h\,,\tilde t=t/r_h\,,\gamma=L/r_h$ and define the adimensional tortoise coordinate $y_*(y)=-\int_y^{\infty} d\tilde{y}/f(\tilde y)$, where we have chosen $y_*(\infty)=0$. The Schr\"{o}dinger equation for the wave function reads
\be -\fft{d^2}{dy_*^2}\psi(y_*)+V_k(y_*) \psi(y_*)=\omega^2 \psi(y_*) \,,\ee
where both the potential $V_k$ and the frequency $\omega$ are dimensionless, normalized by the horizon radius. One has
\be V_k(y_*)=f(y)\Big[\fft{d^2-1}{4\gamma^2}+\fft{4k^2+d^2-4d+11}{4y^2}+\fft{(d-1)^2\big(1+\ft{1}{\gamma^2} \big)}{4y^d} \Big] \,,\ee
where for AdS-Schwarzschild black holes
\be f(y)=1-\fft{1+\ft{1}{\gamma^2}}{y^{d-2}}+\fft{y^2}{\gamma^2} \,.\ee
\begin{figure}
  \centering
  \includegraphics[width=210pt]{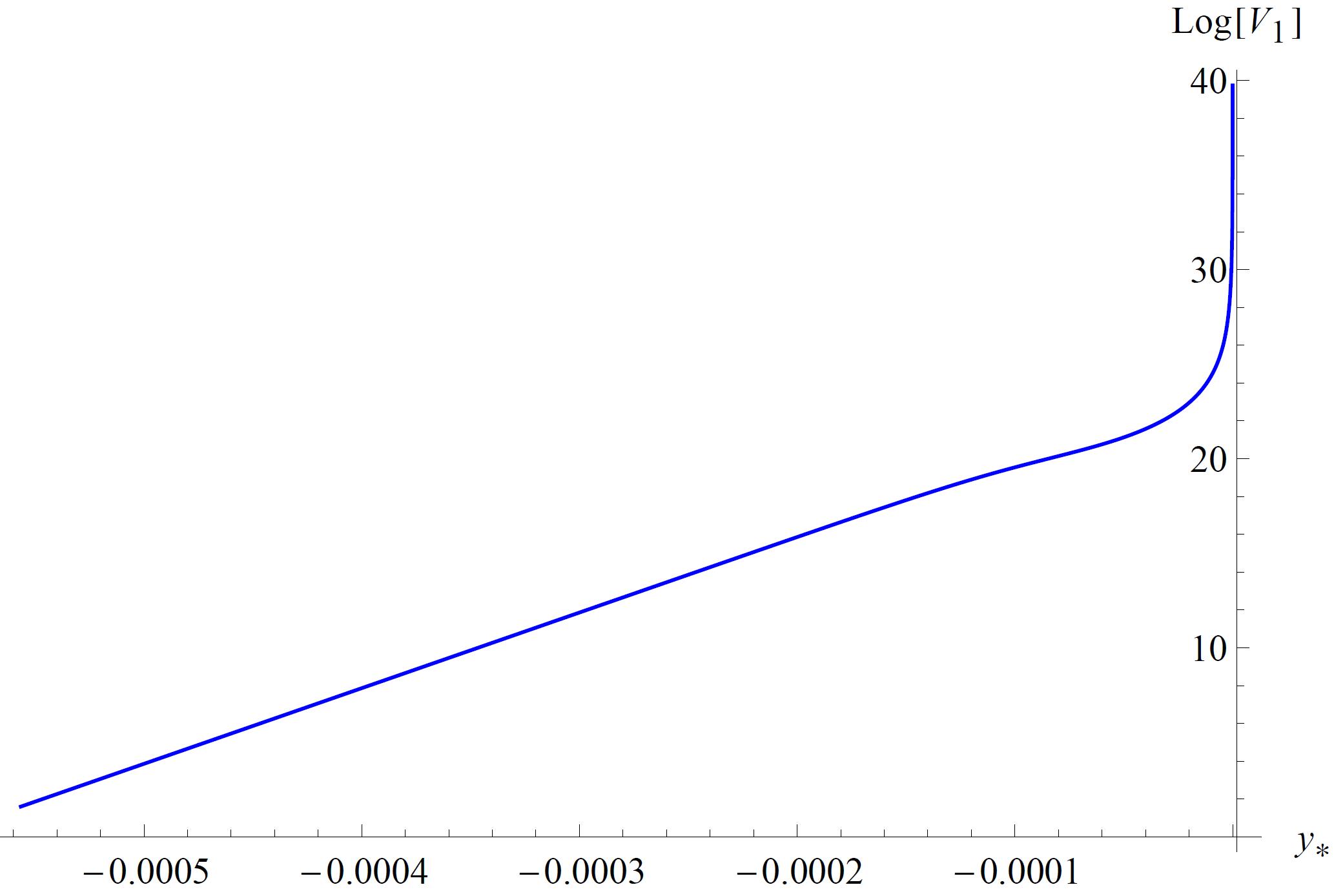}
  \includegraphics[width=210pt]{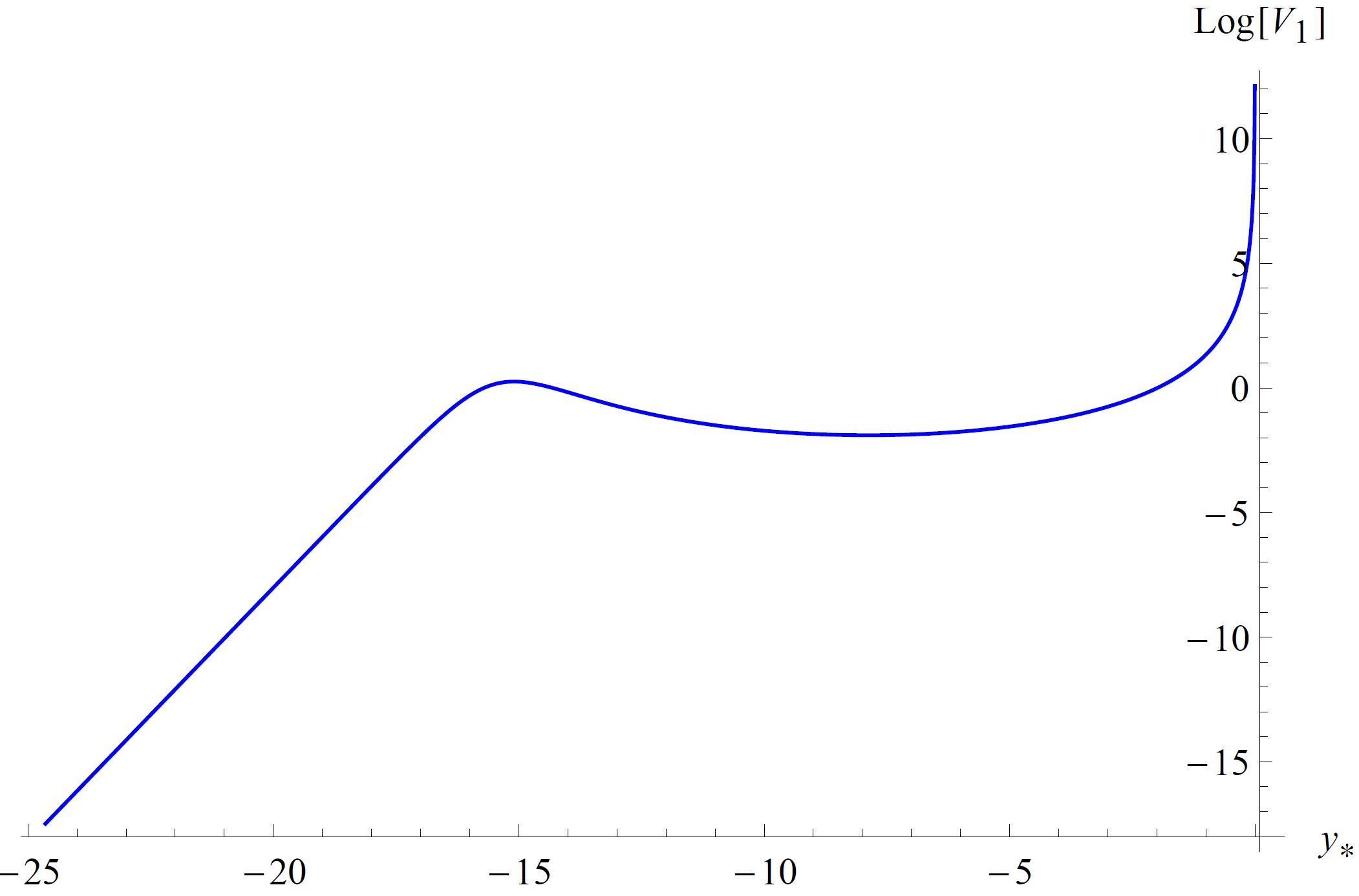}\\
  \caption{The potential $V_1$ for $d=4$ dimensional black holes with angular momentum $\ell=1$. The left panel is for large black holes with $\gamma=0.01$ whereas the right panel is for small black holes with $\gamma=10$. }\label{potential}
\end{figure}
In $d=4$ dimension, the tortoise coordinate can be solved as
\be y_*=\ft{\gamma^2}{2\big( \gamma^2+2 \big)}\Big[ \log{\big(\ft{y-1}{y+1} \big)}+2\sqrt{\gamma^2+1}\,\,\mm{arctan}\Big( \ft{y}{\sqrt{\gamma^2+1}}\Big) -\pi\sqrt{\gamma^2+1}\,\Big]\,.\ee
 Close to the event horizon $y\rightarrow 1$, $y_*\rightarrow -\infty$, the potential vanishes exponentially as $V_k(y_*)\sim e^{2\kappa y_*}$, where $\kappa=1+2/\gamma^2$ is the surface gravity of black holes, whereas at asymptotic infinity $y\rightarrow \infty\,,y_*\rightarrow 0$, the potential diverges in a power-law $V_k(y)\sim 15/4y_*^2$. This is universal to both small and large black holes. However, for large black holes $\gamma< 1$, there is only one meta-stable mode while for small black holes $\gamma> 1$, there exists an overtones for the quasi bound mode due to existence of the potential barrier outside the event horizon, see Fig. \ref{potential}.

 The quasinormal modes of small black holes were carefully studied in \cite{Clarkson:2005mg} for AdS-Schwarzschild black holes with a Einstein static brane. However, in our case, the ETW brane moves dynamically. It is unclear how to impose a proper boundary condition on the brane $y_*=y_{b*}$ in this case. Inspired by \cite{Antonini:2019qkt}, we adopt adiabatic approximation:  taking the ETW brane as quasi-static with respect to the time scale of the gravitational perturbation. Precisely speaking, if the quasinormal mode frequency is given by $\omega=\bar \omega+i\Gamma/2$, then the oscillation time $t_o=1/\bar{\omega}$ of the perturbation should be much smaller than the ``Hubble time" of the brane $T_H=y(t)/y'(t)$, namely
\be t_o\ll T_H=\fft{y\,\phi(y)}{f(y)\sqrt{\phi^2(y)-f(y)}} \,,\ee
where we have used the Lorential version of (\ref{firstint}). We will first compute the quasinormal mode frequency and then test the above condition in the final.

For a static brane with $y=y_b$, the dual normal vector is given by $ n_\mu=(dy)_\mu/\sqrt{f(y)}$ and $e^\mu_a=\delta^\mu_a$. Evaluation of the linearized extrinsic curvature, one finds from the mixed boundary condition (\ref{mixbc})
\be\label{mixedbdy} \partial_{y_*}\psi(y_*)\Big|_{y_*=y_*^b}=\fft{(d-1)f(y)}{2y}\,\psi(y_*) \,.\ee
This is exactly the same as the boundary condition imposed in \cite{Antonini:2019qkt}. However, derivation of the condition in \cite{Antonini:2019qkt} further required that $ y'^2/f^2(y)\ll 1$, which guarantees that interpretation of gravitons for a bulk observer is equivalent to that of a comoving observer on the brane. However, in our case, we do not need it any longer.

As usual, we impose ingoing boundary condition on the event horizon
\be \psi(y_*)|_{y\rightarrow 1}\sim e^{i\omega y_*} \,.\ee
The quasinormal mode frequencies are generally discrete since we impose two boundary conditions.

\subsection{Numerical results}
\begin{figure}
  \centering
  \includegraphics[width=140pt]{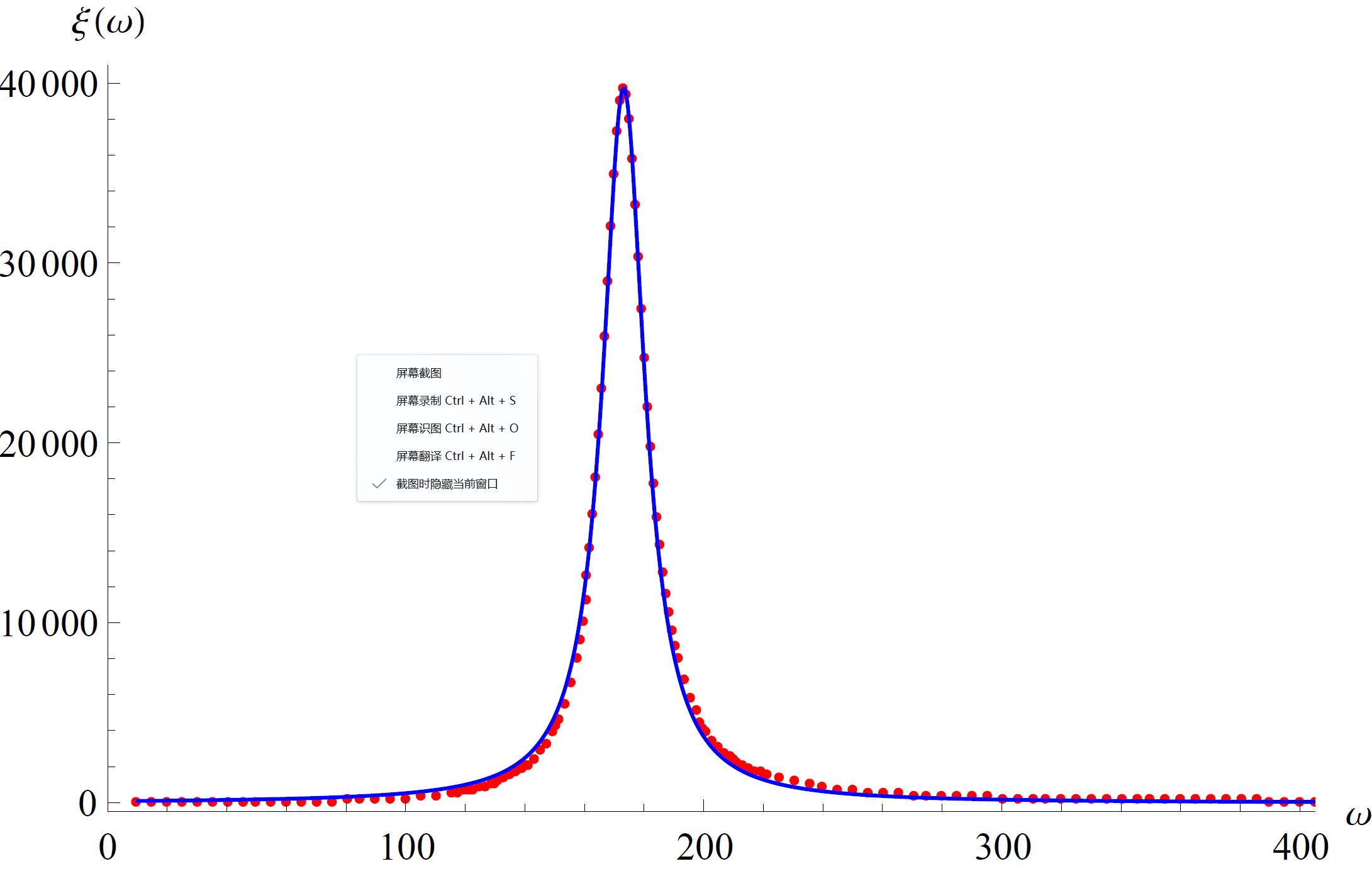}
  \includegraphics[width=140pt]{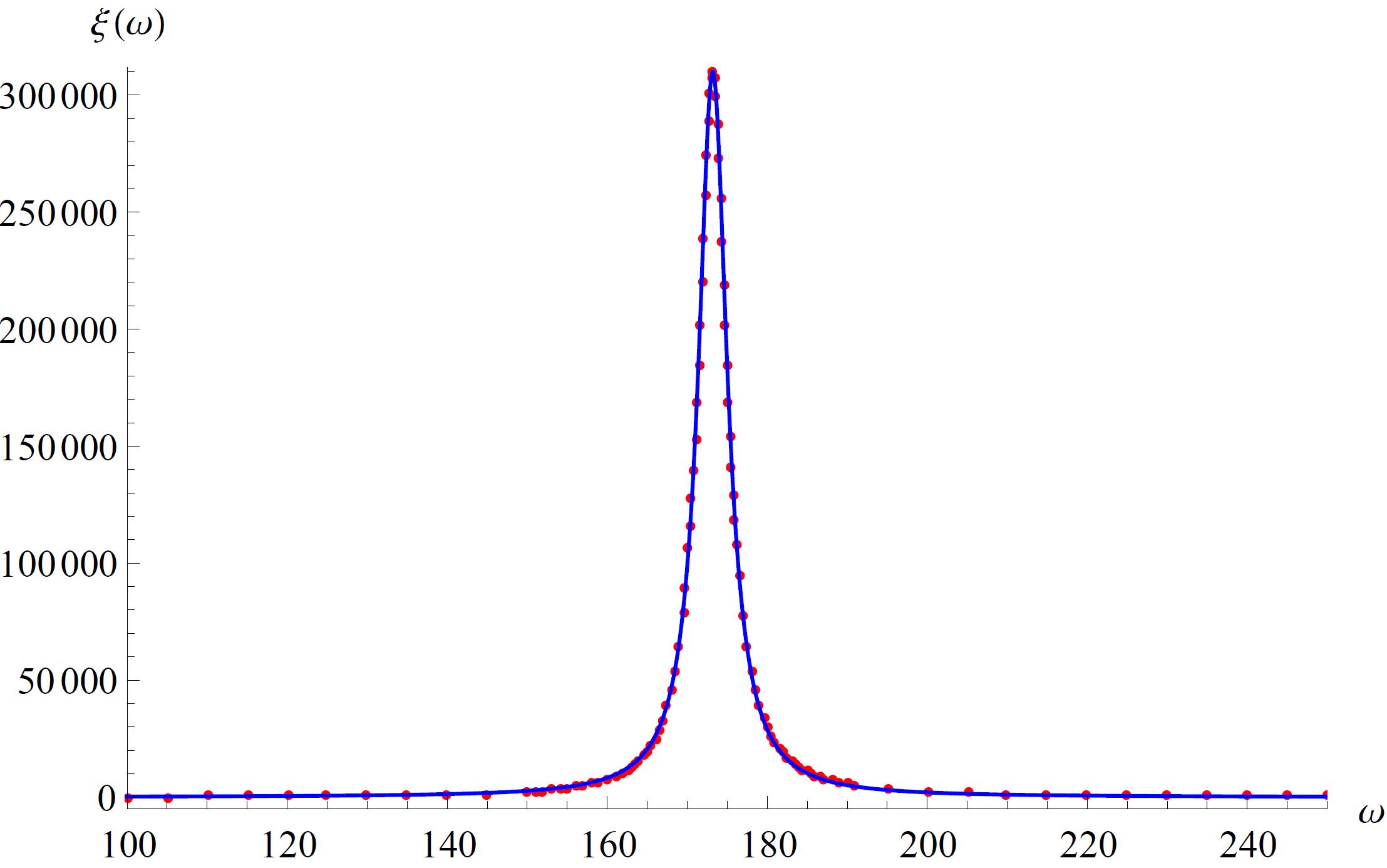}
  \includegraphics[width=140pt]{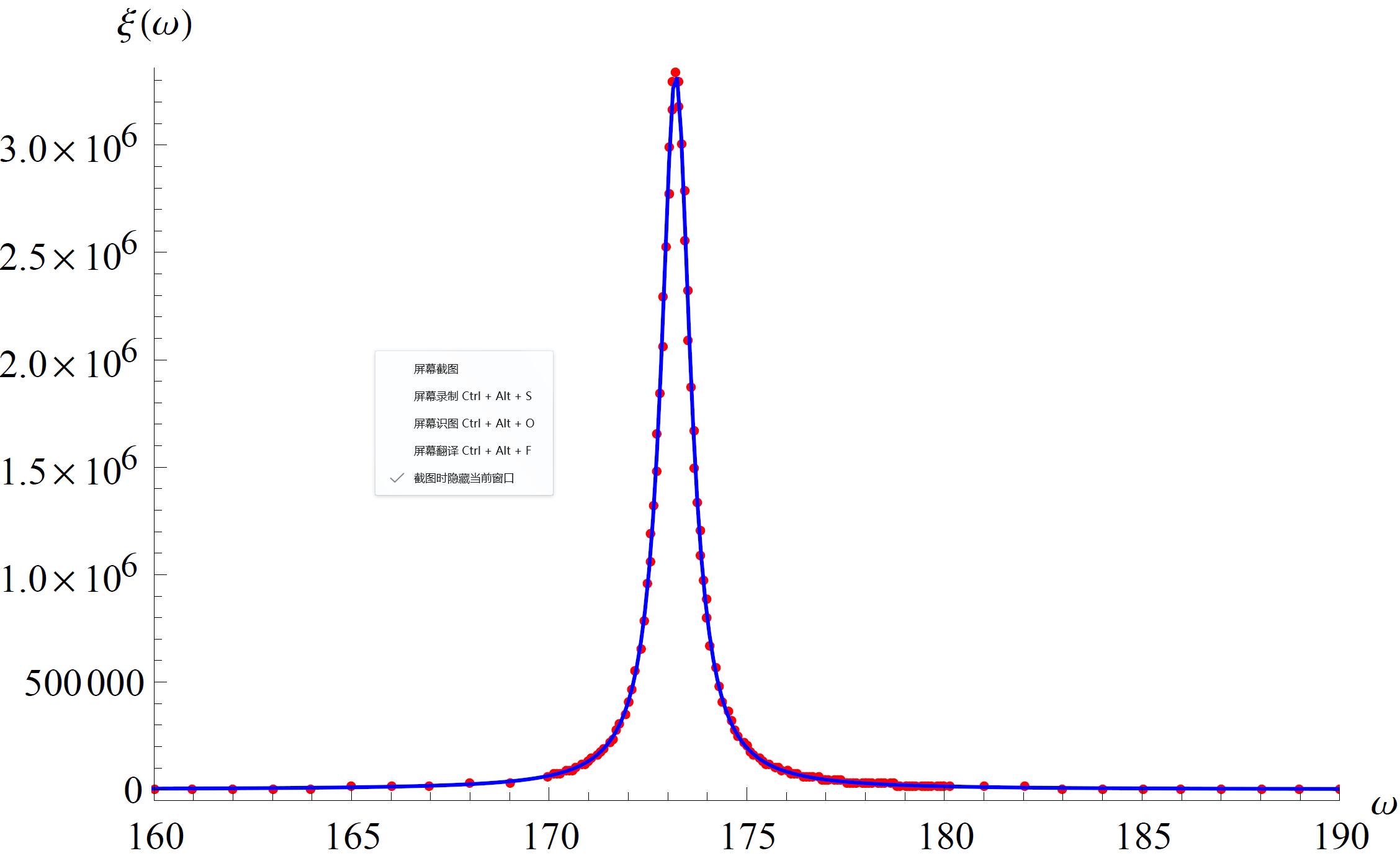}
  \caption{The squared trapping coefficient $\xi(\omega)$ for $d=4$ dimensional black holes with $\gamma=0.01$ and angular momentum $\ell=1$. One has $\omega_{GR}=173.205$. Left panel: $y_b=34.12$. $\bar\omega=173.143\,,\Gamma=17.181$. Middle panel: $y_b=67.69$. $\bar\omega=173.194\,,\Gamma=4.366$. Right panel: $y_b=149.41$. $\bar\omega=173.202\,,\Gamma=0.896$. As the brane stays farther from the black hole, the peak becomes higher and the width becomes narrower. In the limit $y_b\rightarrow \infty$, $\bar\omega\rightarrow \omega_{GR}$ and $\Gamma\rightarrow 0$, the mode reduces to the RS II bound mode.}\label{ga01}
\end{figure}
\begin{table}[h]
\centering
\begin{tabular}{|c|c|c|c|c|}
  \hline
  $y_b$ &          $t_o$           &    $t_d$        &             $T_H$          &  $\lambda(y_b)/\lambda_{tot}$    \\ \hline
    60  & $4.997601\cdot 10^{-6}$  &    0.1406596    &  $4.549750\cdot 10^{-5}$   &   0.7689702                               \\ \hline
    70  & $4.997576\cdot 10^{-6}$  &    0.1914732    &  $5.679376\cdot 10^{-5}$   &   0.6770077                               \\\hline
    80  & $4.997564\cdot 10^{-6}$  &    0.2482983    &  $7.362785\cdot 10^{-5}$   &   0.5603646                              \\ \hline
    90  & $4.997551\cdot 10^{-6}$  &    0.3154222    &  $1.085232\cdot 10^{-4}$   &   0.4007155                               \\ \hline
    99  & $4.997541\cdot 10^{-6}$  &    0.3901243    &  $3.526517\cdot 10^{-4}$   &   0.1277716                               \\ \hline
  99.9  & $4.997538\cdot 10^{-6}$  &    0.3900052    &  $1.117740\cdot 10^{-3}$   &   0.0404218                             \\ \hline
    100 & $4.997537\cdot 10^{-6}$  &    0.3908119    &           $\infty$         &       0                              \\ \hline
\end{tabular}
\caption{Time scales comparison for $d=4$ dimensional black holes with $\gamma=0.01$ and angular momentum $\ell=2000$. We have set $L=1\,,T=0.9999\,,q=10000$ so that the maximum radius of the brane is $y_b^{max}=100$. The second and third columns report the oscillation time $t_o=1/\bar{\omega}$ and decay time $t_d=2/\Gamma$ (or the lift time) of the quasi bound mode. The last column gives the ratio of the proper time $\lambda(y_b)$ needed for the brane to expand from $y_b$ to $y_b^{max}$ to the total proper time needed to complete the trajectory of expansion.}
\label{timescale}
\end{table}
We are interested in the large black hole case $\gamma\ll 1$, which has only one meta-stable mode. We adopt trapping coefficient method \cite{Clarkson:2005mg} to extract the frequency. For self-consistency, we review the method in the Appendix \ref{appendixb}. It turns out that the real and imaginary parts of the quasi bound mode can be read off from the squared trapping coefficient $\xi(\omega)$, which takes the form of
\be \xi(\omega)=R^2(\omega)\fft{ \,\Gamma_n^2}{4(\omega-\bar{\omega}_n)^2+\Gamma_n^2} \,,\ee
where $R(\omega)$ is a slowly varying function of $\omega$. The real part of the frequency of the quasi bound mode corresponds to the Lorentzian (Breit-Wigner) peak while half of its imaginary part $\Gamma/2$ gives rise to half-width of the peak.

In Fig. \ref{ga01}, we show the squared trapping coefficient for four dimensional black holes with $\gamma=0.01$ and angular momentum $\ell=1$. As the location of the brane $y_b$ increases, the Breit-Wiger peak becomes narrower and higher, implying that localization of gravity becomes more efficient for a larger radius brane. On the other hand, fixing $y_b$ and $\gamma$, we find that for larger angular momentum $\ell$, the gravity localization becomes more efficient as well. This is in accordance with a effective description of gravity on the brane locally since gravitons with larger angular momentum $\ell$ probe shorter distance scales. The shorter the spatial scale of gravitational perturbation is, the better the graviton mode is bound on the brane. Our numerical results are consistent with \cite{Clarkson:2005mg}: in the limit $y_b\rightarrow \infty$, $\bar\omega\rightarrow \omega_{GR} $, $\Gamma\rightarrow 0$, the quasi bound mode reduces to the RS II bound mode when the brane approaches to asymptotic infinity. In the Appendix C, we study the limiting behavior of the quasi bound mode half-analytically using the large dimensional limit. The result again supports our numerics.

In Table \ref{timescale}, we compare time scales for a large black hole $\gamma=0.01$ in $d=4$ dimensions. For a given brane tension and dark charge, the adiabatic approximation is valid very well near the inversion point of the brane trajectory irrespective of the angular momentum. For a larger angular momentum, for example $\ell=2000$, it holds also for a significant portion in the trajectory of the brane. For example, if one accepts $t_o\sim T_H/10$ as the threshold for the validity of the approximation, then it holds for more than $70\%$ in the brane trajectory. The larger the angular momentum is, the more portion of the brane trajectory the approximation is valid to. We also observe that the decay time (or lift time) $t_d=2/\Gamma$ of the quasi bound mode is much larger than the Hubble time. This implies that our analysis will break down before the quasi bound mode leaks into the bulk.

\section{Conclusion}

In this paper, we study a holographic model of braneworld cosmology proposed in \cite{Antonini:2019qkt}. The main new contributions we made in this work is proving that the model can be extended to general black hole backgrounds while the results in \cite{Antonini:2019qkt} are limited to charged black holes.

We focus on the simplest setup of AdS-Schwarzschild black holes with a pure tension end-of-the-world (ETW) brane. By imposing a mixed boundary condition on the brane, we find that for a given tension, the motion of the brane is characterized by an integration constant $q$, referred to as {\it dark charge}. Tuning the dark charge, we show that in Euclidean signature the parameters allowing a large radius brane far from the black hole horizon are compatible with those admitting a positive preparation time. The latter is a least condition for sensible Euclidean solutions, which are dual to well-defined CFT states. Furthermore, by numerically evaluating the frequency of the quasi bound mode, we show that gravity is locally localized on the brane for spatial and times scales smaller than the cosmological ones. Therefore, sensible Euclidean solutions and localization of gravity on the brane can accommodate with each other in our setup.

Our analysis can be easily generalised to different black hole backgrounds. Since the dark charge plays an essential role in a consistent holographic description of braneworld cosmology, it is of great importance to explore its physical meaning. There are two hints: \\
$\bullet$ In Euclidean signature, the dark charge $q$ contributes to a part of the stress tensor on the past boundary of the CFT. This is similar to black hole mass. However, it never contributes to the bulk stress tensor.\\
$\bullet$ In real time, physical effects of $q$ was encoded in the motion of the brane, which corresponds to evolution of FLRW cosmology. By deriving the Friedmann equation, we find that the dark charge gives an exra contribution to the energy density in FLRW cosmology. Again, this is similar to the black hole mass. \\

Based on these observations, we argue that if the black hole mass is interpreted as visible matter in our universe, the dark charge may be identified with dark matter in cosmology.  However, this neither gives any hint on the microstructures of dark matter nor explains its gravitational effects in galactic scales. Despite that the argument is far from a consistent story, it still provides a new possibility to study dark matter in holography.

\section*{Acknowledgments}
Z.Y. Fan was supported in part by the National Natural Science Foundations of China with Grant No. 11805041 and No. 11873025.

\appendix

\section{ More discussions on mixed boundary conditions }\label{mixbdy}

In general, one can impose a mixed boundary condition on the ETW brane as
\be C^{ab}\Big[ K_{ab}-\big(K-(d-1)T \big)h_{ab} \Big]=0\,,  \label{genemix}\ee
where $C_{ab}=\lambda_1 h_{ab}+\lambda_2 K_{ab}+\lambda_3 R_{ab}+\cdots $. It was argued in \cite{Miao:2017gyt,Chu:2017aab} that this general choice of boundary condition does not produce correct boundary Weyl anomaly for general shapes of the boundary $P$. However, this does not effect our discussions since in our case $P$ is a past boundary so that its spatial curvature and the boundary Weyl anomaly are trivial.

We first work in Euclidean signature. The unit dual normal vector of the brane is still given by (\ref{norm}), namely
\be n_\mu=N(-r'\,,1\,,\mathbf{0})\,,\qquad N=\sqrt{\fft{f(r)}{f^2(r)+(r')^2}} \,.\ee
By letting $N\equiv \varphi(r)/f(r)$, one finds
\be \fft{dr}{d\tau}=\pm \fft{f(r)}{\varphi(r)}\sqrt{f(r)-\varphi(r)^2}\,,\ee
where the function $\varphi(r)$ should be solved from the boundary condition. We are interested in the case of
\be C_{ab}=\alpha T\, h_{ab}+K_{ab} \,,\ee
where $\alpha$ is an adimensional constant. Without loss of generality, we take $0\leq \alpha< +\infty$, where the limit $\alpha\rightarrow +\infty$ corresponds to the simplest case $C_{ab}=h_{ab}$ studied in the main text.
\begin{figure}\label{fig6}
  \centering
  \includegraphics[width=210pt]{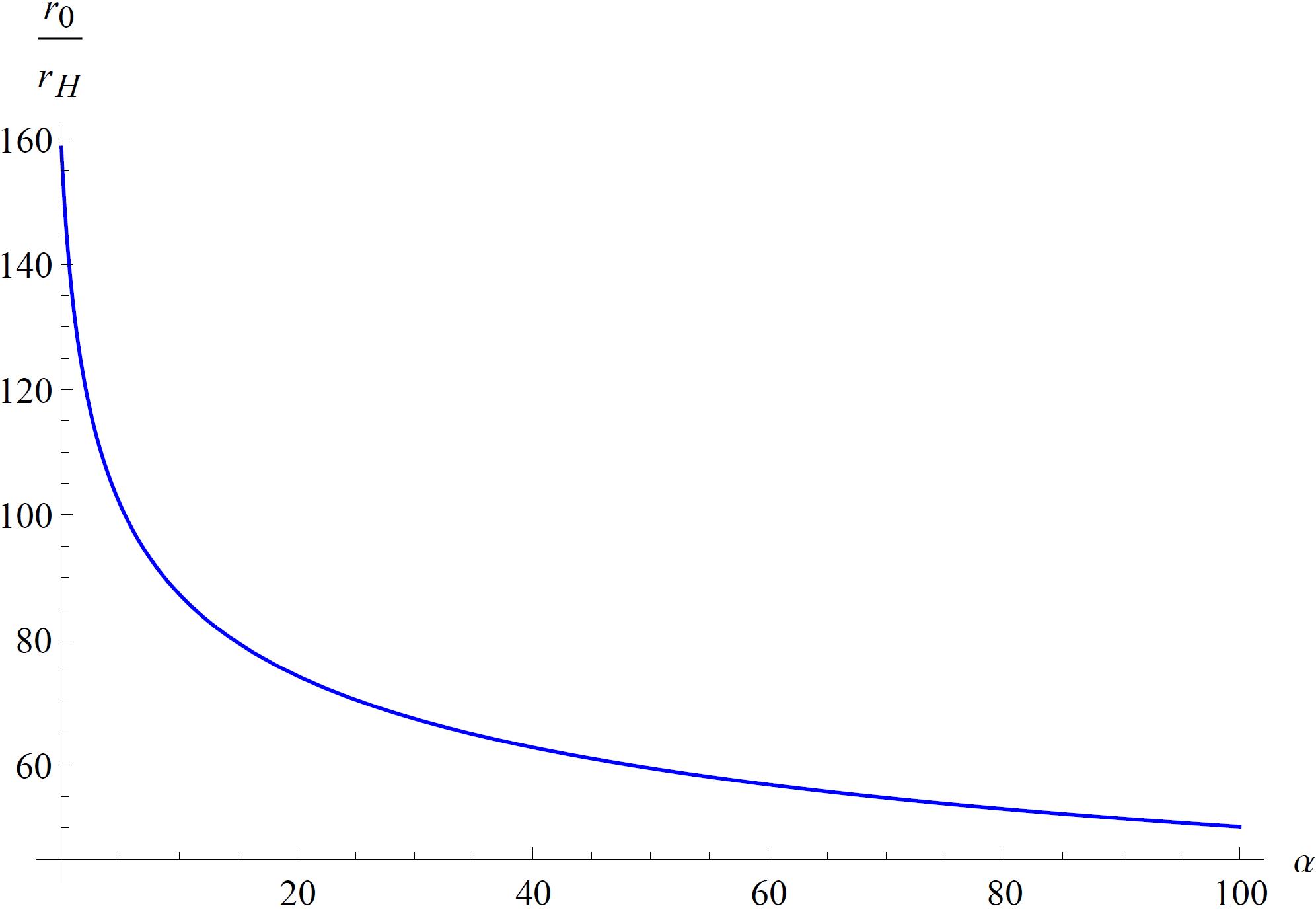}
  \includegraphics[width=210pt]{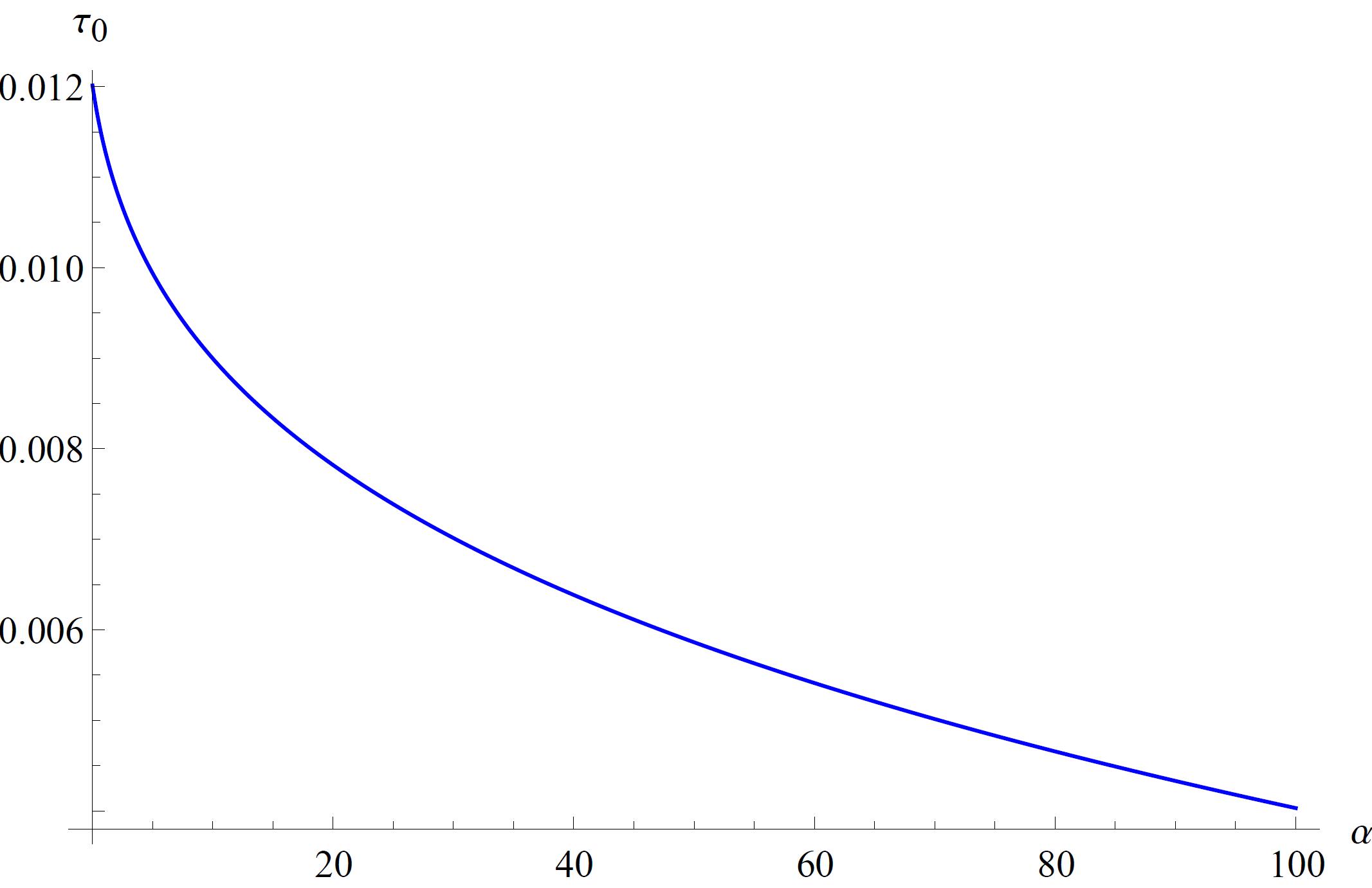}
  \caption{Left panel: the minimum radius for a $d=4$ dimensional black hole with $r_H=100\,,T=0.9\,,q=8\times 10^7$. Right panel: the preparation time for the same black hole with $T=0.999\,,q=2\times 10^6$. Both $r_0$ and $\tau_0$ are decreasing functions of $\alpha$.   }\label{alpha}
\end{figure}
It follows that
\bea
0&=&\sqrt{g}\,\big( \alpha T\, h^{ab}+K^{ab} \big)\Big[ K_{ab}-\big(K-(d-1)T \big)h_{ab} \Big] \nn\\
&=&-(d-1)\partial_r \Big[ r^{d-2}\big(\varphi(r)-T r \big)\big(\varphi(r)+\alpha T\,r \big) \Big]\,.
\eea
Again, we obtain a first integral
\be r^{d-2}\big(\varphi(r)-T r \big)\big(\varphi(r)+\alpha T\,r \big)\equiv q^{d-2} \,,\label{phiphi}\ee
where $q$ is an integration constant, still referred to as {\it dark charge}. We deduce
\be \varphi(r)=\fft12\Big(\sqrt{T_+^2r^2+\ft{q^{d-2}}{r^{d-2}}} +T_- r\Big) \,,\ee
where $T_\pm=T(1\pm \alpha)$. The brane tension is bounded above as $T<1/L$ since $\varphi(r)\rightarrow T r$ in the asymptotic limit. Notice that the above solution cannot reduce to (\ref{firstint}) in the limit $\alpha\rightarrow +\infty$ (but one can rediscover the solution directly from (\ref{phiphi}) despite that the dark charge cannot be identified with one another).
However, it turns out that for a given $\alpha$, basic features of the above solution, such as the minimum radius $r_0$ of the ETW brane and the preparation time $\tau_0$ of the boundary theory as functions of the brane tension $T$ or the dark charge $q$ are qualitatively similar to those established in sec.\ref{sec31}. Moreover, linearizing the mixed boundary condition (\ref{genemix}) around a Schwarzschild black hole leads to the same boundary condition (\ref{mixedbdy}) on the brane for the quasi bound modes of gravitons (in adiabatic approximation). Therefore, under the boundary condition (\ref{genemix}), a positive preparation time in the boundary CFT is again compatible with locally localization of gravity on the ETW brane. In Fig. 6, we show the minimum radius $r_0$ and the preparation time $\tau_0$ as a function of $\alpha$. Interestingly, they both decrease as $\alpha$ increases. It implies that a smaller $\alpha$ is better for realizing braneworld cosmology.

In real time, the Friedmann equation is given by
\be\label{Friedmanngene} \Big( \fft{\dot{r}}{r}\Big)^2=\Big(\ft{T_+^2+T_-^2}{4}-1/L^2\Big)-\fft{1}{r^2}+\fft{2\mu+\ft14 q^{d-2} }{r^d}+\fft{T_-}{2}\,\sqrt{T_+^2+\ft{q^{d-2}}{r^{d}}}\,.\ee
 Again it describes a radiation dominated universe with a negative cosmological constant $3\big( T^2-1/L^2 \big)$ and energy density proportional to the black hole mass plus the contribution from the dark charge. In particular, for $\alpha=1$, the Friedmann equation greatly simplifies to
\be \Big( \fft{\dot{r}}{r}\Big)^2=\Big(T^2-1/L^2\Big)-\fft{1}{r^2}+\fft{2\mu+\ft14 q^{d-2} }{r^d}\,,\ee
where the last unconventional term in (\ref{Friedmanngene}) disappears. In this case, the dark charge plays completely the same role of black hole mass in cosmology.

\section{ Action comparison}\label{actioncompare}
For a given ETW brane with a fixed tension $T$, a dark charge $q$ and a preparation time $\tau_0$, there will be different black hole solutions $r_H$, including the pure AdS. We shall compute the Euclidean action to establish which phase has least action and hence is dominant in the path integral. It is worth emphasizing that for a given $q\neq 0$, there exists nonzero energy flux passing through the ETW brane. While this energy is invisible in the bulk, it plays an indispensable role in our discussions. The action that we would like to compare is the one for different phases with this same ``dark energy". This defines a certain thermodynamic ensemble.

We work in Euclidean signature and focus on the metric of the form
\be\label{Schlike}  ds^2=f(r)d\tau^2+\fft{dr^2}{f(r)}+r^2 d\Omega^2_{d-1} \,.\ee
The induced metric on the ETW brane is given by
\be ds^2_{\mm{ETW}}=\Big( f(r)\big(\fft{d\tau}{dr}\big)^2+\fft{1}{f(r)}\Big)dr^2+r^2 d\Omega^2_{d-1} \,.\ee
The total Euclidean action reads
\bea I&=&-\fft{1}{16\pi G}\int_N d^{d+1}x\,\sqrt{g}\,\big( R-2\Lambda\big)-\fft{1}{8\pi G}\int_{\mm{ETW}} d^dy\,\sqrt{h}\,\Big( K-(d-1) T\Big) \nn\\
&&-\fft{1}{8\pi G}\int_{M} d^dy\,\sqrt{\gamma}\,K-\fft{1}{8\pi G}\int_P d^{d-1}x\,\sqrt{\sigma}\,\theta\,,\eea
where $M$ is the asymptotic boundary at infinity and $P=M\bigcap \mm{ETW}$ stands for the corners at asymptotic infinity.
For the Schwarzschild like metric (\ref{Schlike}), one always has $R-2\Lambda=-2d L^{-2}$. On the ETW brane, $K=d T$ owing to the boundary condition. Moreover, $\sqrt{g}=r^{d-1}$ and
\be \sqrt{h}=r^{d-1}\sqrt{f(r)\big(\fft{d\tau}{dr}\big)^2+\fft{1}{f(r)}}=\fft{r^{d-1}}{\sqrt{f(r)-\varphi^2(r)}}=\pm r^{d-1}\fft{f(r)}{\varphi(r)}\fft{d\tau}{dr} \,,\ee
where we have used the equation of motion for the brane
\be \fft{d\tau}{dr}=\pm \fft{\varphi(r)}{f(r)\sqrt{f(r)-\varphi^2(r)}} \,.\ee
In addition, for $r=\mm{cons}$ hypersurface, the dual normal vector is given by $\tilde{n}_{\mu}=\fft{(dr)_\mu}{\sqrt{f(r)}}$ and $\sqrt{\gamma}=\sqrt{f(r)}\,r^{d-1}$. The extrinsic curvature is evaluated as
\be K\Big|_{r=\mm{cons}}=\fft{f'(r)}{2\sqrt{f(r)}}+\fft{(d-1)\sqrt{f(r)}}{r} \,.\ee
For the corners at asymptotic infinity, a simple calculation gives $\theta=\arccos{T}$, which does not depend on bulk interior geometries.

In the black hole phase, one has on the brane trajectory
\bea\tau(r)=\int_{r_0}^r d\hat{r}\, \fft{\varphi(\hat r)}{f(\hat r)\sqrt{f(\hat r)-\varphi^2(\hat r)}}  \,.
\eea
Hence, the preparation time of the dual CFT is given by
\be \tau_0=\fft{\beta}{2}-\int_{r_0}^\infty dr\, \fft{\varphi( r)}{f( r)\sqrt{f( r)-\varphi^2( r)}}  \,,\ee
where for Schwarzschild black holes,
\be f(r)=\fft{r^2}{L^2}+1-\fft{r_H^{d-2}}{r^{d-2}}\Big( \fft{r_H^2}{L^2}+1 \Big)\,,\quad  \beta= \fft{4\pi r_H L^2}{d r_H^2+(d-2)L^2}  \,.\ee
For pure AdS, the result depends on whether $q$ is vanishing or nonvanishing. When $q=0$, the configuration has two disconnected branches bounded by the ETW brane.
This case has been studied carefully in \cite{Cooper:2018cmb}. The conclusion is for given $(T\,,\tau_0)$, the largest black hole phase is dominant.

We shall focus on a nonvanishing $q\neq 0$. In this case, the brane has a minimum radius even in pure AdS. The situation is similar to the black hole phase except that the Euclidean time periodicity $\beta$ is chosen by hand (precisely speaking, by matching asymptotic geometries of pure AdS with that of black holes)
\be \tau_0=\fft{\hat{\beta}}{2}-\int_{\hat{r}_0}^\infty dr\, \fft{\varphi( r)}{f_{AdS}(r)\sqrt{f_{AdS}(r)-\varphi^2( r)}}  \,,\ee
where $f_{AdS}=r^2/L^2+1$. In the following, we will set $L=1$ for simplicity.

\textbf{Black hole phase}: For the black hole phase, the bulk action is given by the total action for the Euclidean black hole (up to some UV cut-off $r_{max}$) minus the action for the excised part. This gives
\be \fft{\Omega_{d-1}}{8\pi G }\Big[ \beta \int_{r_H}^{r_{max}}dr\,dr^{d-1}-\int_{r_0}^{r_{max}}dr\,dr^{d-1}\,2\tau(r) \Big] \,.\ee
The brane action is given by
\be -\fft{\Omega_{d-1}}{8\pi G}\int_{r_0}^{r_{max}}dr\,2Tr^{d-1}\fft{f(r)}{\varphi(r)}\fft{d\tau}{dr} \,.\ee
These two parts lead to
\bea
&&\fft{\Omega_{d-1}}{4\pi G }\Big[ \fft{\beta}{2}r^d\Big|_{r_H}^{r_{max}}-\int_{r_0}^{r_{max}}dr\,\Big( dr^{d-1}\tau(r)+T r^{d-1}\fft{f(r)}{\varphi(r)}\fft{d\tau}{dr} \Big)\Big] \nn\\
&&=\fft{\Omega_{d-1}}{4\pi G }\Big[ \fft{\beta}{2}r^d\Big|_{r_H}^{r_{max}}-r^d\tau(r)\Big|^{r_{max}}_{r_0}-\int_{r_0}^{r_{max}}dr\,\Big( \ft{T}{r} -\ft{\varphi(r)}{f(r)}\Big)r^d\fft{f(r)}{\varphi(r)}\fft{d\tau}{dr}\Big]\,,
\eea
where in the second line we have used $\big( r^d\tau(r)\big)'=dr^{d-1}\tau(r)+r^d d\tau/dr$.

Combined with the GHY surface term evaluated at asymptotic infinity and the corner terms, we deduce
\bea\label{bhaction} I_{BH}&=&\fft{\Omega_{d-1}}{4\pi G L^2}\Big[  \tau_0\, r^{d-1}\big(r-\ft{f'(r)}{2}-\ft{(d-1)f(r)}{r} \big)_{r=r_{max}}-\theta\,r_{max}^{d-1}\nn\\
&&\qquad\qquad-\fft12 \beta r_H^d-\int_{r_0}^{r_{max}}dr\,\Big( \ft{T}{r} -\ft{\varphi(r)}{f(r)}\Big)\fft{r^d}{\sqrt{f(r)-\varphi^2(r)}}\Big] \,.\eea

\textbf{Pure AdS}: Since $q\neq 0$, the ETW brane in pure AdS will also have a connected configuration similar to the black hole case. The action for Euclidean black holes (\ref{bhaction}) is valid to pure AdS as well, with all the quantities replaced by those of thermal AdS $r_H\rightarrow 0\,,r_{\max}\rightarrow \hat{r}_{max}\,,r_0\rightarrow \hat{r}_0\,,f(r)\rightarrow f_{AdS}(r)\,,\tau_0\rightarrow \hat{\tau}_0\,,\beta\rightarrow\hat{\beta}\,,\theta\rightarrow \hat{\theta}$. One has
\bea I_{AdS}&=&\fft{\Omega_{d-1}}{4\pi G }\Big[\hat{\tau}_0\, r^{d-1}\big(r-\ft{f_{AdS}'(r)}{2}-\ft{(d-1)f_{AdS}(r)}{r} \big)_{r=\hat{r}_{max}}\nn\\
&&\qquad\qquad-\hat\theta\,\hat{r}_{max}^{d-1}-\int_{\hat{r}_0}^{\hat{r}_{max}}dr\,\Big( \ft{T}{r} -\ft{\varphi(r)}{f_{AdS}(r)}\Big)\fft{r^{d}}{\sqrt{f_{AdS}(r)-\varphi^2(r)}}\Big]\,.
\eea
Notice that the Euclidean periodicity $\hat{\beta}$ can be chosen arbitrarily so that the preparation time $\hat{\tau}_0$ matches with the black hole case at asymptotic infinity
$\tau_0=\hat{\tau}_0|_{\hat{r}_{max}\rightarrow \infty}$.

 Before explicitly evaluating the action difference, we shall point out that for Schwarzschild black holes, the Euclidean periodicity is up bounded as $\beta\leq \beta_c=\fft{2\pi L}{\sqrt{d(d-2)}}$, where the equality corresponds to a critical horizon $r_{H,c}=\sqrt{\fft{d-2}{d}}
L$ and a temperature at which the Hawking-Page transition occurs. For a given Euclidean periodicity $\beta$, in general there exists two phases: a smaller black hole with $r_{H}<r_{H,c}$ and a larger black hole with $r_H>r_{H,c}$. It follows that the preparation time $\tau_0$ shares a similar behavior to $\beta$ for given $T\,,q$: it is up bounded as $\tau_0\leq \tau_0(r_{H\,,c})$ and for a given $\tau_0$ below the maximum value, there exists a smaller black hole phase with $r_{H}<r_{H,c}$ and a larger black hole phase with $r_H>r_{H,c}$, respectively. For our purpose, we are interested in the larger black hole phase. The preparation time $\tau_0$ and $\beta$ are in a one-to-one map with the horizon radius. In other words, fixing $T\,,q\,,\beta$ is equivalent to fixing $T\,,q\,,\tau_0$ for the regions of parameters of interests.

To evaluate the action difference, we need match the geometries in the asymptotic region for both cases. First, we choose the asymptotic cutoffs $r_{max}\,,\hat{r}_{max}$ such that they match with the cutoff $z=\epsilon$ in Fefferman-Graham coordinates
\be \fft{Ldz}{z}=\fft{dr}{\sqrt{f(r)}} \,,\ee
where the integration constant in the above equation is chosen such that $r\sim 1/\epsilon$ to leading order. It is easily found that for pure AdS in general dimension
\be \hat{r}_{max}=\fft{1}{\epsilon}-\fft{\epsilon}{4} \,,\ee
and for $d=4$ dimensional black holes
\be r_{max}=\fft{1}{\epsilon}-\fft{\epsilon}{4}+\fft{1}{8}r_H^2(r_H^2+1)\epsilon^3+\fft{1}{32}r_H^2(r_H^2+1)\epsilon^5+O(\epsilon^7) \,.\ee
Moreover, we need match the proper Euclidean preparation times in the asymptotic region\footnote{For the matching conditions we imposed, the GHY surface term evaluated at asymptotic infinity has nontrivial contributions in the action difference. However, if one extends the relation $\tau_0=\hat{\tau}_0$ at asymptotic infinity to any finite cutoff, the GHY surface term will no longer contribute to the action difference. The paper \cite{Antonini:2019qkt} has a minor mistake in this sense. }
\be 2\tau_0\sqrt{f(r_{max})}=2\hat{\tau}_0 \sqrt{f_{AdS}(\hat{r}_{max})}  \,.\ee
The lowest order in $\epsilon$ gives (for Schwarzschild black holes)
\be \hat{\tau}_0(\epsilon)=\tau_0\big(1-\ft{(d-1)\mu}{d}\,\epsilon^d \big) \,.\ee

Combining all the results above, we deduce the action difference
\bea\label{actiondiff} \delta I&\equiv &\fft{4\pi G}{\Omega_{d-1}}\big( I_{AdS}-I_{BH}\big)  \nn\\
&=&\tau_0\Big((d-1)\big( r_{max}^{d}+r_{max}^{d-2}-\hat{r}_{max}^{d}-\hat{r}_{max}^{d-2}+\ft{(d-1)\mu}{d}\big)-d\mu \Big)\nn\\
&&+\big(\theta r_{max}^{d-1}-\hat\theta\hat{r}_{max}^{d-1} \big)+\fft12 \beta r_H^d\nn\\
&&-\int_{\hat{r}_0}^{\hat{r}_{max}}dr\,\Big( \ft{T}{r} -\ft{\varphi(r)}{f_{AdS}(r)}\Big)\fft{r^d}{\sqrt{f_{AdS}(r)-\varphi^2(r)}}\nn\\
&&+\int_{r_0}^{r_{max}}dr\,\Big( \ft{T}{r} -\ft{\varphi(r)}{f_{BH}(r)}\Big)\fft{r^d}{\sqrt{f_{BH}(r)-\varphi^2(r)}}
\,.\eea
\begin{figure}\label{actioncptq}
  \centering
  \includegraphics[width=210pt]{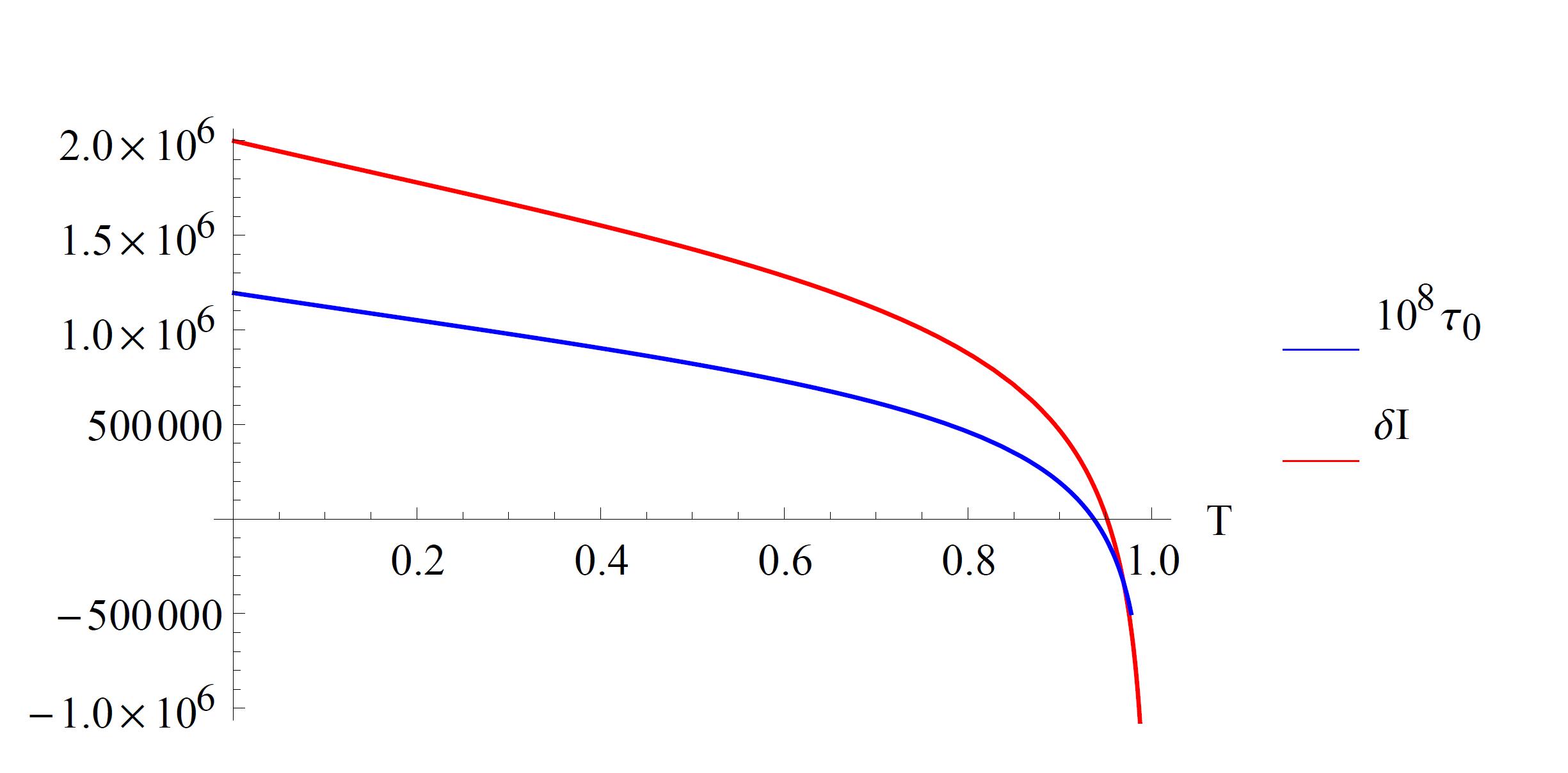}
  \includegraphics[width=210pt]{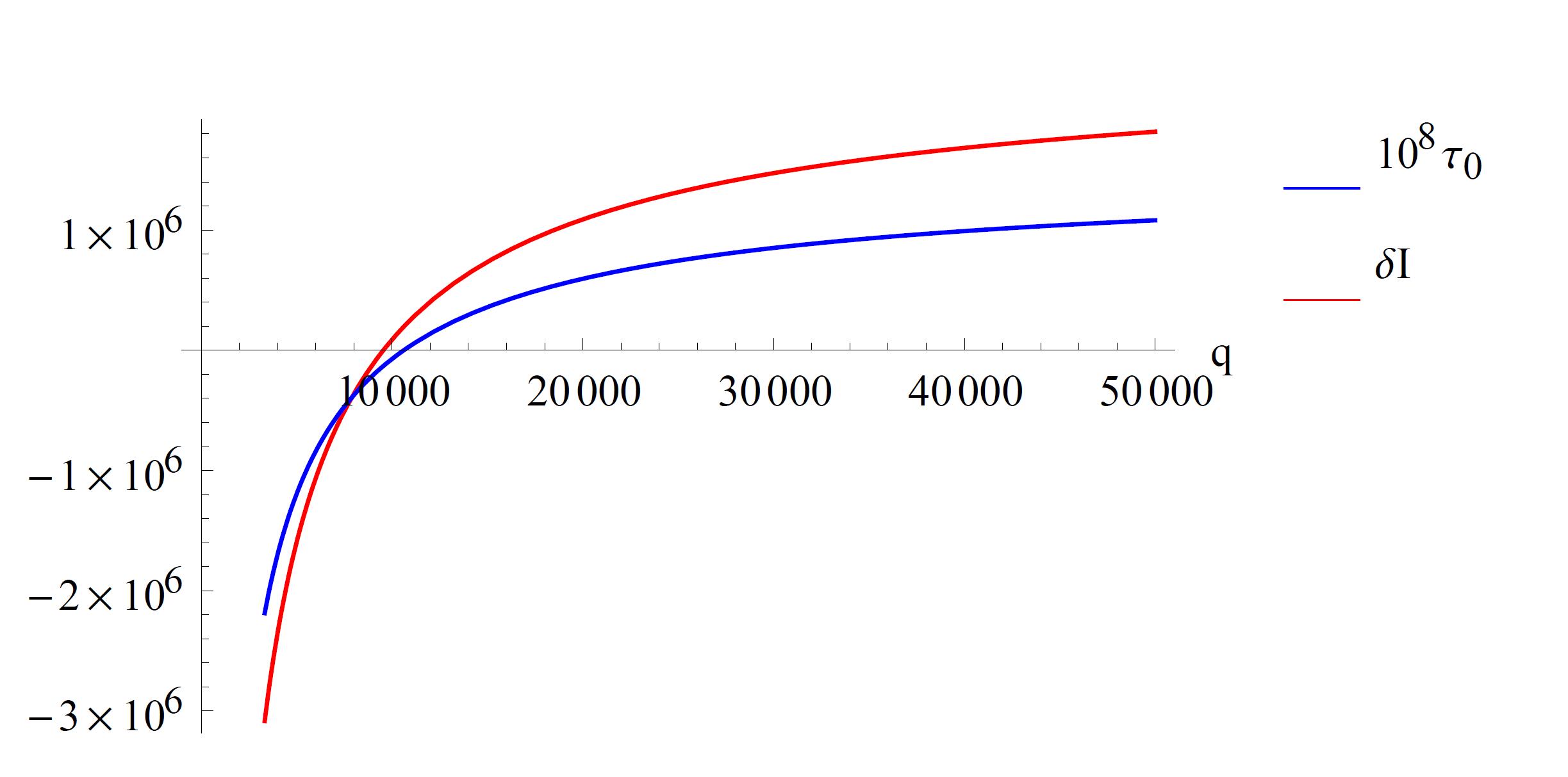}\\
  \caption{The action difference $\delta I$ and the preparation time $\tau_0$ for a large black hole with $r_H=100$. Left panel: For a fixed charge $q$ (we choose $q=500$), the preparation time is positive for a brane tension not too close to the critical value $T_{crit}=1/L$ and the action difference is always positive for any $\tau_0\geq 0$. It implies that for a sensible Euclidean solution, the black hole phase is dominant in the path integral. Right panel: For a near critical tension (we choose $T=0.99999$), the preparation time becomes positive for a sufficiently large charge $q$ and the action difference is already positive before this happens, again implying that the black hole phase is dominant for a sensible Euclidean solution. }
\end{figure}
It should be emphasized that the GHY surface term evaluated at asymptotic infinity gives nonvanishing contributions.


In the $d=4$ dimension, one has
\be \delta I=\fft{\pi r_H^5}{2r_H^2+1}+\fft{5}{8} r_H^2(r_H^2+1)\tau_0-\fft{T \big( r_0 -\hat{r}_0\big)}{\sqrt{1-T^2}}+I_{4,AdS}+I_{4\,BH} \,,\ee
where
\bea
&&I_{4,AdS}=-\int_{\hat{r}_0}^{\infty}dr\,\Big[\big( \ft{T}{r} -\ft{\varphi(r)}{f_{AdS}(r)}\big)\fft{r^4}{\sqrt{f_{AdS}(r)-\varphi^2(r)}}-\fft{T }{\sqrt{1-T^2}} \Big]\,,\nn\\
&&I_{4,BH}=\int_{r_0}^{\infty}dr\,\Big[\big( \ft{T}{r} -\ft{\varphi(r)}{f_{BH}(r)}\big)\fft{r^4}{\sqrt{f_{BH}(r)-\varphi^2(r)}}-\fft{T }{\sqrt{1-T^2}} \Big]\,.
\eea
Numerical results for $d=4$ dimensions are shown in Fig. 7. For a given $q\neq 0$, the preparation time is positive for a brane tension not too close to the critical value $T_{crit}=1/L$. In this case, the action difference is always positive for any $\tau_0\geq 0$, see the left panel. On the other hand, for a near critical tension (for example $T=0.99999$), the preparation time can still be positive for a sufficiently large $q$. In the right panel, one finds that the action difference becomes already positive before this happens. These results strongly imply that the large black hole phase is dominant for a sensible Euclidean solution even in the presence of nonzero energy flux passing through the ETW brane.

\section{Trapping coefficient method}\label{appendixb}

In the near horizon region, the potential $V_{k\,,\omega}$ vanishes and the wave function $\psi_{k\,,\omega}$ will take the form of a plane wave
\be \psi_{k\,,\omega}=\fft{A_h}{2}e^{-i\delta(\omega)}\Big( e^{-i\omega y_*}+S(\omega) e^{i\omega y_*} \Big) \,,\ee
where $S(\omega)=e^{2i\delta(\omega)}$ is the scattering matrix and $\delta(\omega)$ is the scattering phase shift. Since the solution is infalling on the horizon, the scattering matrix should have some poles (at complex frequencies) so that the second term in the bracket is dominant. Recall that we impose an extra boundary condition on the brane, these poles will be discrete, giving rise to a discrete set of quasi-bound modes (labelled by $\omega_n$). The leading Laurent expansion for the scattering matrix is given by \cite{Taylor}
\be S(\omega)=e^{2i\delta_0(\omega)}\fft{\omega-\omega_n^*}{\omega-\omega_n} \,,\ee
where $\delta_0(\omega)$ is a slowly varying function of $\omega$. Using the definition $\omega_n=\bar{\omega}_n+i\Gamma_n/2$, one finds
\be \delta(\omega)=\delta_0(\omega)+\mm{arcsin}\Big[ \fft{\Gamma_n}{\sqrt{4(\omega-\bar{\omega}_n)^2+\Gamma_n^2}}\Big] \,.\ee

On the other hand, in the asymptotic region, the wave function is just a slowly-varying function of $\omega$. If its magnitude at the brane is denoted by $A_b$, one can write \cite{Clarkson:2005mg}
\bea
\psi_{k\,,\omega}=\left\{\begin{array}{ll}
A_h\, \mm{Re}\big[ e^{i\delta(\omega)} e^{i\omega y_*} \big]\,,\quad y_*\rightarrow -\infty\\
\\
A_b\equiv A_h R(\omega)\,\mm{Re}\big[ e^{i\delta(\omega)} e^{i\theta(\omega)} \big]=-A_h R(\omega)\sin\big(\delta(\omega)+\theta(\omega)-\ft{\pi}{2} \big)\,,\quad y_*=y_{b*}\,,
\end{array}
\right.
\eea
where $R(\omega)\,,\theta(\omega)$ are slowly-varying functions of $\omega$. The trapping coefficient is defined as
\be \eta(\omega)=\fft{A_b}{A_h}=-R(\omega)\sin\big(\delta(\omega)+\theta(\omega)-\ft{\pi}{2} \big)\,.\ee
In practical calculations, one can choose a proper normalization for the wave function such that $\psi_{k\,,\omega}(y_{b*})=A_b=1$. This does not change the trapping coefficient but is convenient for numerical calculations. Naively, if the quasi bound mode is locally localized on the brane due to the presence of an attractive delta potential, we may expect the squared trapping coefficient has a peak at the frequency of the quasi bound mode.

If the relation
\be\label{condition} \delta_0(\omega)+\theta(\omega)=\fft{\pi}{2}\,,\fft{3\pi}{2} \,,\ee
holds, the squared trapping coefficient will be given by
\be \xi(\omega)=\eta^2(\omega)=R^2(\omega)\,\fft{\Gamma_n^2}{4(\omega-\bar{\omega}_n)^2+\Gamma_n^2} \,.\label{xi}\ee
This is the well-known Lorentzian (Breit-Wigner) distribution. The peak is centered at the real part of the frequency of the quasi-bound mode whereas its half-width is equal to the half of the imaginary part of the frequency $\Gamma_n/2$. Of course, if the condition (\ref{condition}) is not satisfied, the relation (\ref{xi}) is invalid and the squared trapping coefficient will take a more complicated form. However, it was shown in \cite{Clarkson:2005mg} that (\ref{condition}) is a good approximation and our numerical results confirm this again. As was pointed out in \cite{Clarkson:2005mg}, to obtain the Lorentzian distribution, one generally needs subtract from the plot of $\xi(\omega)$ a baseline function, which accounts for the slow variation of $\delta_0(\omega)\,,\theta(\omega)$ with frequency. However, this is not necessary in our situation since we are interested in the large black hole case, in which only the zero mode is present.

It turns out that the trapping coefficient method is a straightforward algorithm to extract the real and imaginary parts of the quasi bound mode:
(1) Numerically solving the Schr\"{o}dinger equation for a range of real frequencies, given the boundary condition on the brane, together with the normalization $\psi_{k\,,\omega}(y_{b*})=1$; (2) Find numerically the amplitude of the wave function $A_h$ in the near horizon region. The square of its inverse $1/A_h^2$ gives the squared trapping coefficient $\xi(\omega)$; (3) Plot $\xi(\omega)$ and subtract a baseline function; (4) Fit the numerical data with Breit-Wigner distrubtion and read off the real and imaginary parts of the frequency of the quasi bound mode.

\section{Limiting behavior of the quasi bound mode}

In general, quasinormal modes of black holes can only be numerically studied. However, in the large dimensional limit ($D=d+1$), the gravitational field of a black hole is well localized in a region near the horizon so that the frequency of quasinormal modes can be half-analytically extracted
\cite{Emparan:2014cia,Emparan:2014aba,Emparan:2015rva}. The quasinormal modes can be classified into two types according to their frequency: decoupling modes and non-decoupling modes \cite{Emparan:2014cia,Emparan:2014aba}. The first refers to the modes with smaller frequency, which are trapped in the near horizon region whereas the latter stands for the larger frequency modes which can penetrate the gravitational potential barrier and escape to asymptotic infinity with certain probability. For example, for asymptotically flat black holes, decoupling modes has $\omega\sim O(1)$ whereas nondecoupling modes has $\omega\sim O(D)$.

In our case, we are interested in the limiting behavior of the quasi bound mode for a far brane close to the asymptotic infinity. The quasi bound mode of course corresponds to the non-decoupling modes. Without confusion, we send the brane to asymptotic infinity and will not distinguish the two in the following. In addition, we focus on large black holes which has $\gamma\ll 1$ and the frequency of the quasi bound mode should obey  $\mm{Re}\,\omega<\omega_{GR}$, where the two should be in the same order of $D$.

\begin{figure}
  \centering
  \includegraphics[width=270pt]{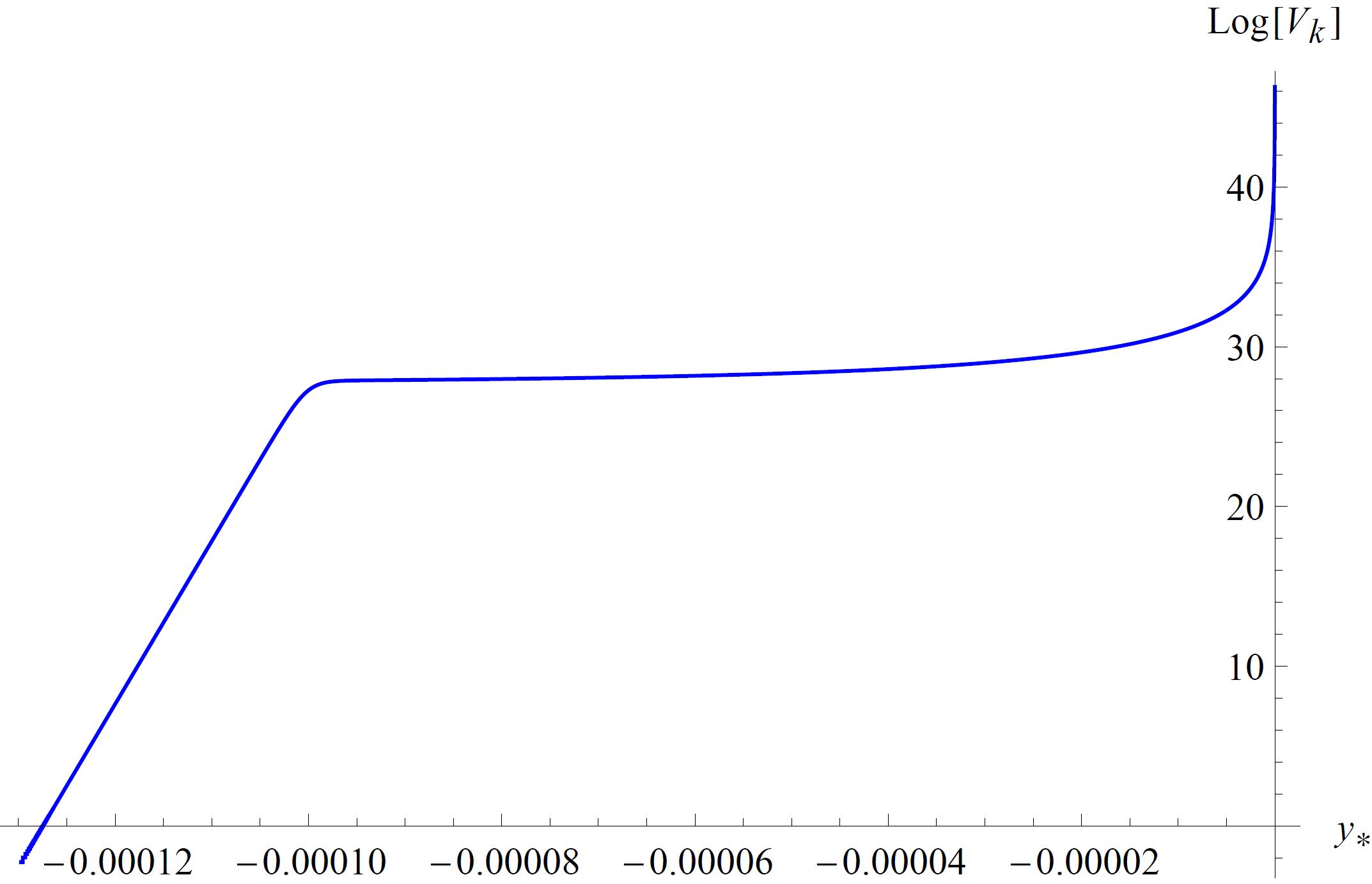}
  \caption{The gravitational potential for a large black hole in $n=100$ dimensions with $\gamma=1/100$ and angular momentum $\ell=100^2$.}\label{VLBHD}
\end{figure}

According to these considerations, we choose $\omega\sim O(n^3)\,,\gamma\sim O(1/n)\,,\ell\sim O(n^2)$, where $n=D-3$. It is convenient to introduce $\hat{\omega}=\omega/n^3\,,\hat{\gamma}=
n\gamma\,,\hat{\ell}=\ell/n^2$ and a new radial coordinate $R=y^n$. It turns out that the gravitational potential approaches to a plateau for $R\gg 1$ before arriving at asymptotic infinity, see Fig. \ref{VLBHD}
\be V_k(R)\Big|_{R\gg 1}\longrightarrow n^6 \omega_c^2\,,\quad \omega_c=\sqrt{\hat{\omega}_{GR}^2+\fft{1}{4\hat{\gamma}^4} } \,,\ee
where $\hat{\omega}_{GR}\equiv \omega_{GR}/n^3  =\hat{\ell}/\hat{\gamma}$. Clearly, smaller frequency modes with $\omega\ll n^3 \omega_c$ will be trapped inside the potential barrier, staying in the near horizon region since their wave functions will vanish exponentially in $n$ outside the barrier. These are the decoupling modes of AdS-Schwarzschild black holes, which are studied carefully in \cite{Emparan:2015rva}. The modes that we are interested in are the nondecoupling ones which have higher frequency $\omega\sim n^3\omega_c$. For our purpose, we would like to take $\omega_{GR}\sim n^3 \omega_c$ so that $\mm{Re}\,\omega\sim \omega_{GR}$. This demands $\hat{\ell}\gg 1/\hat{\gamma}$ and hence $\hat{\omega}_{GR}\simeq \omega_c$.

Here it should be emphasized that the large black hole considered above features a resonant cavity in the bulk. In this case, there will exist an overtones for each $\ell$ for the quasi bound mode. This is significantly different from the $d=4$ dimensional case. The frequency of the quasi bound mode can be extracted by matching the near horizon solutions and the far zone solutions in the overlap region $1/n\ll y-1\ll 1$ (or $1\ll R\ll e^n$).

\subsection{Near zone solutions}

In the near horizon region with $ y-1\ll 1$ (or $R\ll e^n$), the metric function $f(R)$ and the potential $V_k(R)$ simplify to
\bea
&&f(R)=\fft{n^2}{\hat{\gamma}^2}\Big(1-\fft{1}{R} \Big)\,,\nn\\
&&V_k(R)=f(R)n^4\Big( \hat{\gamma}^2 \omega_c^2+\fft{1}{ 4\hat{\gamma}^2 R} \Big)\,,
\eea
where we keep only the leading order in the large $D$ limit. The ingoing boundary condition implemented on the horizon is $\psi(R)= \big(R-1 \big)^{i\hat{\omega}\hat{\gamma}^2}\phi(R)$, where $\phi(R)$ is a regular function at $R=1$. Solving the master equation yields
\be \psi(R)=\big(R-1 \big)^{i\hat{\omega}\hat{\gamma}^2}\sqrt{R}\,\,{}_2F_1\big(\alpha_+\,,\alpha_-\,,1\,;R \big) \,,\ee
where
\be \alpha_\pm=\fft12+i\hat{\omega}\hat{\gamma}^2\pm\hat{\gamma}^2 \sqrt{\omega_c^2-\hat{\omega}^2}  \,.\ee
In the overlap region $R\gg 1$, the solution asymptotes to
\be \psi(R)\longrightarrow A_+(\omega)\,R^{\hat{\gamma}^2\sqrt{\omega_c^2-\hat{\omega}^2}}+A_-(\omega)\,R^{-\hat{\gamma}^2\sqrt{\omega_c^2-\hat{\omega}^2}}\,,\ee
where explicit expressions for the coefficients $A_\pm (\omega)$ are unimportant except that their ratio is of order unity
\be \Big| \fft{A_+(\omega)}{A_-(\omega)}\Big|\sim O(1) \,.\ee

\begin{figure}
  \centering
  \includegraphics[width=270pt]{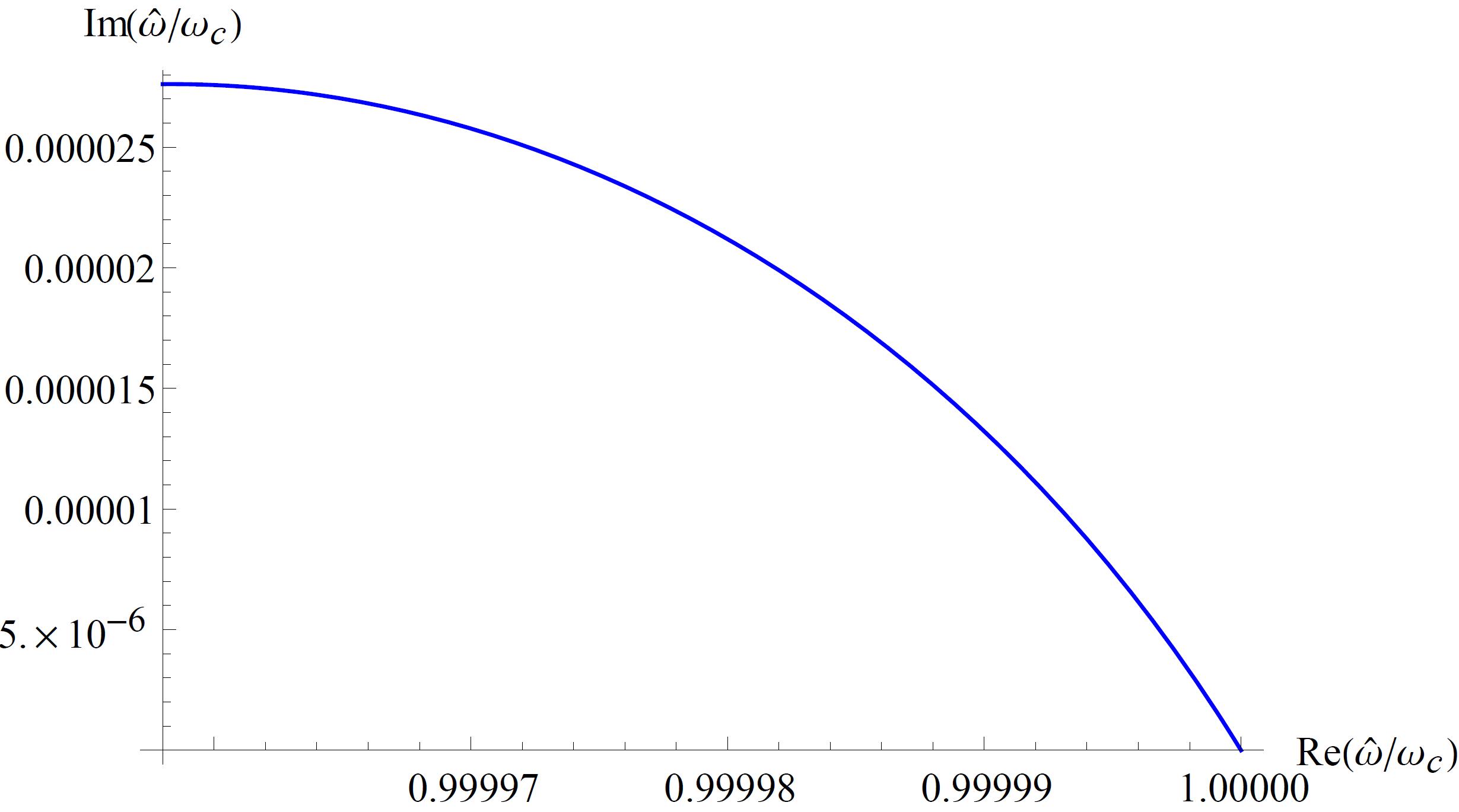}
  \caption{The solution to (\ref{solug}) gives a continuous spectrum for the quasi bound mode. The spectrum should be discretized into separate overtones when subleading order corrections in $1/n$ expansion are included.}
\label{freD}\end{figure}

\subsection{Far zone solutions}
In the far zone where $y-1\gg 1/n$, the term $1/y^n$ is exponentially small in $n$ and hence can be dropped in the metric function. In addition, the spatial curvature is also considerably small compared to the leading term $y^2/\gamma^2$ in asymptotic infinity so that $f(y)= n^2 y^2/\hat{\gamma}^2$ to leading order. The gravitational potential simplifies to
\be V_k(y)=n^6\Big(\omega_c^2+\fft{y^2-1}{4\hat{\gamma}^4} \Big) \,.\ee
In the large $D$ limit, the mixed boundary condition (\ref{mixedbdy}) imposed on the ETW brane becomes
\be \psi(y)\Big|_{y\gg 1}\longrightarrow y^{n/2} \,.\ee
The solution to the master equation, which obeys the above condition is given by the Hankel function
\be \psi(y)=y^{-1/2}\,H_{n/2}^{(1)}\Big( \ft{ n\hat{\gamma}^2\sqrt{\hat{\omega}^2-\hat{\omega}_{GR}^2}}{y} \Big) \,.\ee
In the overlap region $1\ll R\ll e^n$, the solution behaves asymptotically as\footnote{The asymptotic behavior of the Hankel function can be found in the appendix in \cite{Emparan:2014aba}. }
\be \psi(R)\longrightarrow A_+(\omega)\,R^{\hat{\gamma}^2\sqrt{\omega_c^2-\hat{\omega}^2}}+A_-(\omega)\,R^{-\hat{\gamma}^2\sqrt{\omega_c^2-\hat{\omega}^2}}\,,\ee
where
\be
\Big| \fft{A_+(\omega)}{A_-(\omega)}\Big|=\mm{exp}\Big[n \,\mm{Re}\,g ( x ) \Big] \,,\quad x=2\hat{\gamma}^2 \sqrt{\hat{\omega}^2-\hat{\omega}_{GR}^2}\,,\ee
where
\be g(x)=\log\Big(\fft{1+\sqrt{1-x^2}}{x} \Big)-\sqrt{1-x^2} \,. \ee
Matching the above far zone solutions to the near zone ones leads to
\be\label{solug} \mm{Re}\,g\Big(2\hat{\gamma}^2 \sqrt{\hat{\omega}^2-\hat{\omega}_{GR}^2}\,\Big)=0 \,.\ee
This gives rise to a continuous spectrum for the quasi bound mode, see Fig. \ref{freD}. However, this is a artifact in the large $D$ limit. Inclusion of subleading order corrections should discretize the spectrum into separate overtones. Nonetheless, this half-analytical result supports our numerical results for the limiting behavior of the quasi bound mode.

\end{document}